# Photometric multi-site campaign on the open cluster NGC 884

## I. Detection of the variable stars

S. Saesen[1,⋆], F. Carrier[1,⋆⋆], A. Pigulski[2], C. Aerts[1,3], G. Handler[4], A. Narwid[2], J. N. Fu[5], C. Zhang[5], X. J. Jiang[6], J. Vanautgaerden[1], G. Kopacki[2], M. Stęślicki[2], B. Acke[1,⋆⋆], E. Poretti[7], K. Uytterhoeven[7,8], C. Gielen[1], R. Østensen[1], W. De Meester[1], M. D. Reed[9], Z. Kołaczkowski[2], G. Michalska[2], E. Schmidt[4], K. Yakut[1,10,11], A. Leitner[4], B. Kalomeni[12], M. Cherix[13], M. Spano[13], S. Prins[1], V. Van Helshoecht[1], W. Zima[1], R. Huygen[1], B. Vandenbussche[1], P. Lenz[4], D. Ladjal[1], E. Puga Antolín[1], T. Verhoelst[1,⋆⋆], J. De Ridder[1], P. Niarchos[14], A. Liakos[14], D. Lorenz[4], S. Dehaes[1], M. Reyniers[1], G. Davignon[1], S.-L. Kim[15], D. H. Kim[15], Y.-J. Lee[15], C.-U. Lee[15], J.-H. Kwon[15], E. Broeders[1], H. Van Winckel[1], E. Vanhollebeke[1], C. Waelkens[1], G. Raskin[1], Y. Blom[1], J. R. Eggen[9], P. Degroote[1], P. Beck[4], J. Puschnig[4], L. Schmitzberger[4], G. A. Gelven[9], B. Steininger[4], J. Blommaert[1], R. Drummond[1], M. Briquet[1,⋆⋆], and J. Debosscher[1]

[1] Instituut voor Sterrenkunde, Katholieke Universiteit Leuven, Celestijnenlaan 200 D, 3001 Leuven, Belgium
e-mail: sophie@ster.kuleuven.be
[2] Instytut Astronomiczny Uniwersytetu Wrocławskiego, Kopernika 11, 51-622 Wrocław, Poland
[3] Department of Astrophysics, Radboud University Nijmegen, POBox 9010, 6500 GL Nijmegen, The Netherlands
[4] Institut für Astronomie, Universität Wien, Türkenschanzstrasse 17, 1180 Wien, Austria
[5] Department of Astronomy, Beijing Normal University, Beijing 100875, China
[6] National Astronomical Observatories, Chinese Academy of Sciences, Beijing 100012, China
[7] INAF - Osservatorio Astronomico di Brera, via Bianchi 46, 23807 Merate, Italy
[8] Laboratoire AIM, CEA/DSM-CNRS-Université Paris Diderot; CEA, IRFU, SAp, centre de Saclay, F-91191, Gif-sur-Yvette, France
[9] Department of Physics, Astronomy, & Materials Science, Missouri State University, 901 S. National, Springfield, MO 65897, USA
[10] Institute of Astronomy, University of Cambridge, Madingley Road, Cambridge CB3 0HA, UK
[11] Department of Astronomy & Space Sciences, Ege University, 35100 Izmir, Turkey
[12] Izmir Institute of Technology, Department of Physics, 35430 Izmir, Turkey
[13] Observatoire de Genève, Université de Genève, Chemin des Maillettes 51, 1290 Sauverny, Switzerland
[14] Department of Astrophysics, Astronomy and Mechanics, National and Kapodistrian University of Athens, Panepistimiopolis, GR 157 84 Zografos, Athens, Greece
[15] Korea Astronomy and Space Science Institute, Daejeon 305-348, South Korea

**ABSTRACT**

*Context.* Recent progress in the seismic interpretation of field $\beta$ Cep stars has resulted in improvements of the physics in the stellar structure and evolution models of massive stars. Further asteroseismic constraints can be obtained from studying ensembles of stars in a young open cluster, which all have similar age, distance and chemical composition.
*Aims.* To improve our comprehension of the $\beta$ Cep stars, we studied the young open cluster NGC 884 to discover new B-type pulsators, besides the two known $\beta$ Cep stars, and other variable stars.
*Methods.* An extensive multi-site campaign was set up to gather accurate CCD photometry time series in four filters ($U$, $B$, $V$, $I$) of a field of NGC 884. Fifteen different instruments collected about 77500 CCD images in 1286 hours. The images were calibrated and reduced to transform the CCD frames into interpretable differential light curves. Various variability indicators and frequency analyses were applied to detect variable stars in the field. Absolute photometry was taken to deduce some general cluster and stellar properties.
*Results.* We achieved an accuracy for the brightest stars of 5.7 mmag in $V$, 6.9 mmag in $B$, 5.0 mmag in $I$ and 5.3 mmag in $U$. The noise level in the amplitude spectra is 50 $\mu$mag in the $V$ band. Our campaign confirms the previously known pulsators, and we report more than one hundred new multi- and mono-periodic B-, A- and F-type stars. Their interpretation in terms of classical instability domains is not straightforward, pointing to imperfections in theoretical instability computations. In addition, we have discovered six new eclipsing binaries and four candidates as well as other irregular variable stars in the observed field.

**Key words.** Galaxy: open cluster and associations: individual: NGC 884 – Techniques: photometric – Stars: variables: general – Stars: oscillations – Binaries: eclipsing

## 1. Introduction

The $\beta$ Cep stars are a homogeneous group of B0-B3 stars whose pulsational behaviour is interpreted in terms of the $\kappa$ mechanism activated by the metal opacity bump. Given that mainly low-degree low-order pressure (p-) and gravity (g-) modes are excited, these stars are good potential targets for in-depth seismic studies of the interior structure of massive stars.

⋆ Aspirant Fellow of the Fund for Scientific Research, Flanders
⋆⋆ Postdoctoral Fellow of the Fund for Scientific Research, Flanders

The best-studied $\beta$ Cep stars are V836 Cen (Aerts et al. 2003), $\nu$ Eridani (Pamyatnykh et al. 2004; Ausseloos et al. 2004; Dziembowski & Pamyatnykh 2008, and references given in these papers), $\theta$ Ophiuchi (Handler et al. 2005; Briquet et al. 2005, 2007), 12 Lacertae (Handler et al. 2006; Dziembowski & Pamyatnykh 2008; Desmet et al. 2009), $\beta$ Canis Majoris (Mazumdar et al. 2006) and $\gamma$ Peg (Handler et al. 2009). The detailed seismic analysis of these selected stars led to several new insights into their internal physics. Non-rigid rotation and core convective overshoot are needed to explain the observed pulsation frequencies. Moreover, for some of the observed modes of $\nu$ Eridani and 12 Lacertae excitation problems were encountered (Daszyńska-Daszkiewicz et al. 2005).

The next challenging step in the asteroseismology of $\beta$ Cep stars is to measure those stars with much higher signal-to-noise from space (e.g., Degroote et al. 2009b) and to study them in clusters. Obviously, we would greatly benefit from the fact that the cluster members have a common origin and formation, implying much tighter constraints when modelling their observed pulsation behaviour. Furthermore, clusters are ideally suited for gathering CCD photometry, providing high-accuracy measurements for thousands of stars simultaneously.

Three clusters were initially selected for this purpose: one southern cluster, NGC 3293, which contains eleven known $\beta$ Cep stars (Balona 1994), and two northern clusters, NGC 6910 and NGC 884, which contain four and two bona fide $\beta$ Cep stars, respectively (Kołaczkowski et al. 2004; Krzesiński & Pigulski 1997). Preliminary results for these multi-site cluster campaigns can be found in Handler et al. (2007, 2008) for NGC 3293, in Pigulski et al. (2007) and Pigulski (2008) for NGC 6910 and in Pigulski et al. (2007) and Saesen et al. (2008, 2009) for NGC 884. In these preliminary reports, the discovery of some new $\beta$ Cep and other variable stars was already announced.

This paper deals with the detailed analysis of NGC 884 and is the first in a series on this subject, presenting our data set and discussing the general variability in this cluster.

## 2. The target cluster NGC 884

NGC 884 ($\chi$ Persei, $\alpha_{2000} = 2^{\rm h}22.3^{\rm m}$, $\delta_{2000} = +57°08'$) is a rich young open cluster located in the Perseus constellation. Together with NGC 869 (h Persei) it forms the Perseus double cluster, which is well documented in the literature.

The possible co-evolution of both clusters puzzled many researchers. Some claim that h and $\chi$ Persei have the same distance modulus and age, others state that h Persei is closer and younger than $\chi$ Persei. For an extensive overview of the photographic, photo-electric and photometric studies on the co-evolution of the two clusters based on the age, reddening and distance moduli, we refer to Southworth et al. (2004a). They conclude that the most recent studies (Keller et al. 2001; Marco & Bernabeu 2001; Capilla & Fabregat 2002; Slesnick et al. 2002) converge towards an identical distance modulus of $11.7\pm0.05$ mag and a log (age/yr) of $7.10\pm0.01$ dex. The average reddening of $\chi$ Persei amounts to $E(B-V) = 0.56 \pm 0.05$.

Slettebak (1968) collected rotational velocities for luminous B giant stars located in the vicinity of the double cluster. He reports that the measured stellar rotation velocities are about 50% higher than for field counterparts, supported by his observation of an unusual high number of Be stars in the clusters. More recent projected rotational velocity measurements of Strom et al.

(2005) confirm that the B stars in h and $\chi$ Persei rotate on average faster than field stars of similar mass and age. More searches for and studies of Be stars in the clusters have been carried out, leading to the detection of 20 Be stars in $\chi$ Persei (see Malchenko & Tarasov 2008, and references therein).

The first variability search in the cluster NGC 884 was conducted by Percy (1972) in the extreme nucleus of the cluster by means of photo-electric measurements. He identified three candidate variable stars on a time scale longer than ten hours, Oo 2088, Oo 2227 and Oo 2262 (Oosterhoff numbers, see Oosterhoff 1937), and one on a shorter time scale of six hours, Oo 2299. Waelkens et al. (1990) set up a photometric campaign spanning eight years. They report that at least half of the brighter stars are variable and that most of them seem to be Be stars. The time sampling of both data sets was not suitable to search for $\beta$ Cep stars. The campaign and analysis by Krzesiński & Pigulski (1997) yielded more detailed results. Based on a photometric CCD search in the central region of NGC 884, they discovered two $\beta$ Cep stars, Oo 2246 and Oo 2299, showing two and one pulsation frequencies, respectively. Furthermore, nine other variables were found: two eclipsing binaries (Oo 2301, Oo 2311), three Be stars (Oo 2088, Oo 2165, Oo 2242), two supergiants (Oo 2227, Oo 2417), possibly one ellipsoidal binary (Oo 2371), and one variable star of unknown nature (Oo 2140). Afterwards, a larger cluster field was studied by Krzesiński (1998) and Krzesiński & Pigulski (2000), each of them leading to one more $\beta$ Cep candidate: Oo 2444 and Oo 2809, but the data sets were insufficient to confirm their pulsational character. Finally, two papers prior to our study are devoted to binaries: Southworth et al. (2004b) investigated the eclipsing binary Oo 2311 in detail and Malchenko (2007) the ellipsoidal variable Oo 2371. Given the occurrence of several variables as well as at least one eclipsing binary, we judged $\chi$ Persei to be well suited for our asteroseismology project.

## 3. Equipment and observations

In 2005, a multi-site campaign was set up to gather differential time-resolved multi-colour CCD photometry of a field of the cluster NGC 884 that contains the two previously known $\beta$ Cep stars. The goal was to collect accurate measurements with a long time base to make the detection of pulsation frequencies at milli-magnitude level possible for a large number of cluster stars of the spectral type B. The observations were taken in different filters to be able to identify their modes.

The entire campaign spanned 800 days spread over three observation seasons. During the first season (August 2005 – March 2006), four telescopes assembled 250 hours of data. The main campaign took place in the second season (July 2006 – March 2007), when nine more telescopes joined the project and gathered 940 hours of measurements. In the third season (July 2007 – October 2007), only the two dedicated telescopes, the 120-cm Mercator telescope at Observatorio del Roque de los Muchachos (ORM) and the 60-cm at Białków Observatory, observed NGC 884 for 100 more hours. In total, an international team consisting of 61 observers used 15 different instruments attached to 13 telescopes to collect almost 77 500 CCD images in the *UBVI* filters and 92 hours of photo-electric data in the *uvby* and $UB_1B_2V_1VG$ filters.

In Table 1, an overview of the different observing sites with their equipment (telescope, instrument characteristics and filters) is given. Almost all sites used CCD cameras, only Observatorio Astronómico Nacional de San Pedro Mártir (OAN-SPM) and ORM made use of photometers. Simultaneous Strömgren *uvby*

**Table 1.** Observing sites and equipment

| ID | Observatory | Country | Longitude | Latitude | Altitude | Telescope | CCD camera/chip | FOV (arcmin²) | Resolution (arcsec/pixel) | Filters |
|---|---|---|---|---|---|---|---|---|---|---|
| 1 | OAN-SPM | Mexico | 117.46°W | 31.04°N | 2776m | 150-cm | (Simultaneous $uvby$ photometer) | – | – | Strömgren $uvby$ |
| 2a | Baker | Missouri, USA | 93.04°W | 37.40°N | 418m | 40-cm | Apogee 7p | 6 × 6 | 0.35 | Johnson $V$ |
| 2b | | | | | | | Roper Sci. 1340B | 12.1 × 11.7 | 0.54 | Johnson $BV$ |
| 3a | ORM | La Palma, Spain | 17.88°W | 28.76°N | 2333m | 120-cm (Mercator) | EEV 40-42 (Merope) | 6.5 × 6.5 | 0.19 | Geneva-Cousins $UBV(I_C)$ |
| 3b | | | | | | | P7 (photometer) | – | – | Geneva $UB_1BB_2V_1VG$ |
| 4 | OFXB | Switzerland | 7.60°E | 46.20°N | 2200m | 60-cm | Apogee 47p | 20 × 20 | 1.15 | Johnson $BVI$ |
| 5 | Michelbach | Austria | 15.76°E | 48.09°N | 636m | 40-cm | Apogee 6 | 26 × 26 | 1.52 | Johnson $BVI$ |
| 6 | Vienna | Austria | 16.34°E | 48.23°N | 241m | 80-cm (vlt) | Apogee 7p | 14.3 × 9.5 | 0.56 | Johnson $BVI$ |
| 7 | Białków | Poland | 16.66°E | 51.47°N | 220m | 60-cm | Andor Tech. iKon DW432 | 12.9 × 11.9 | 0.62 | Johnson-Cousins $BV(I_C)$ |
| 8 | Athens | Greece | 23.78°E | 37.97°N | 250m | 40-cm | SBIG ST-8 | 15 × 10 | 0.59 / 1.77 | Bessell $VI$ |
| 9 | EUO | Turkey | 27.28°E | 38.40°N | 795m | 40-cm | Apogee Alta | 9.2 × 9.2 | 0.54 | Bessell $BV$ |
| 10a | TUG | Turkey | 30.34°E | 36.82°N | 2555m | 150-cm | Fairchild CCD447 (TFOSC) | 13.5 × 13.5 | 0.79 | Bessell $UBV$ |
| 10b | | | | | | | SBIG ST-8 | 13.8 × 9.2 | 0.54 | Bessell $BV$ |
| 11 | Xinglong | China | 117.57°E | 40.40°N | 960m | 50-cm | VersArray 1300B | 22 × 22 | 1.00 | Johnson $V$ |
| 12 | SOAO | South Korea | 128.46°E | 36.93°N | 1378m | 61-cm | SITe | 20.5 × 20.5 | 0.60 | Johnson-Cousins $BV(I_C)$ |

Observatory abbreviations and observers are: **1** – OAN-SPM (Observatorio Astronómico Nacional de San Pedro Mártir) – EP, KU; **2** – Baker – MDR, JRE, GAG; **3** – ORM (Observatorio del Roque de los Muchachos) – SS, JV, BA, CG, RØ, WDM, KY, MC, MSp, SP, VVH, WZ, RH, BV, DLa, EPA, TV, JDR, SD, MR, GD, EB, HVW, EV, CW, GR, YB, PD, JB, RD; **4** – OFXB (Observatoire François-Xavier-Bagnoud) – SS; **5** – Michelbach – ALe, PB, LS; **6** – Vienna – GH, ES, ALe, PL, DLo, BS; **7** – Białków – AP, AN, GM, SZ, XJI; **8** – Athens – PN, ALi; **9** – EUO (Ege University Observatory) – KY, BK; **10** – TUG (Tübitak National Observatory) – KY, BK; **11** – Xinglong – JNF, CZ, XJI; **12** – SOAO (Sobaeksan Optical Astronomy Observatory) – S-LK, DHK, Y-IL, C-UL, J-HK

**Table 2.** Observations summary

| ID | Observatory | Observing time (nights) | Observing time (hours) | Number of frames $U$ | $B$ | $V$ | $I$ | Total | Precision (mmag) $U$ | $B$ | $V$ | $I$ | $\sigma_{amp,V}$ (mmag) | PSF function |
|---|---|---|---|---|---|---|---|---|---|---|---|---|---|---|
| 1 | OAN-SPM | 15 | 63.9 | – | – | – | – | – | – | – | – | – | – | – |
| 2a | Baker | 1 | 1.8 | – | – | 111 | – | 111 | – | – | 6.1 | – | 1.03 | Penny 1 |
| 2b | | 6 | 33.9 | – | 2558 | 2558 | – | 5122 | – | 7.5 | 7.0 | – | 0.25 | Moffat 25 |
| 3a | ORM | 135 | 399.0 | 2969 | 2968 | 2997 | 2932 | 11866 | 3.8/6.6 | 3.6/5.0 | 3.9/5.4 | 4.1/5.4 | 0.18 | Moffat 25 |
| 3b | | 14 | 28.6 | – | – | – | – | – | – | – | – | – | – | – |
| 4 | OFXB | 12 | 64.5 | – | 1107 | 1133 | 1089 | 3329 | – | 5.4 | 5.1 | 7.5 | 0.27 | Moffat 25 |
| 5 | Michelbach | 9 | 18.5 | – | 479 | 476 | 475 | 1430 | – | 7.8 | 4.8 | 5.7 | 0.39 | Gaussian |
| 6 | Vienna | 17 | 90.7 | – | 4301 | 2140 | 3255 | 9696 | – | 5.1 | 10.7 | 4.8 | 0.41 | Moffat 25 |
| 7 | Białków | 71 | 287.7 | – | 2830 | 12438 | 5405 | 20673 | – | 2.6 | 3.0 | 3.8 | 0.05 | Moffat 25 |
| 8 | Athens | 8 | 24.3 | – | – | 967 | 970 | 1937 | – | – | 6.8/5.3 | 6.8/4.8 | 0.35 | Penny 2 / Moffat 15 |
| 9 | EUO | 3 | 13.0 | – | 1244 | 1245 | – | 2489 | – | 5.7 | 6.3 | – | 0.32 | Penny 2 |
| 10a | TUG | 2 | 8.8 | 106 | 101 | 105 | – | 312 | 5.8 | 4.4 | 4.4 | – | 0.76 | Moffat 25 |
| 10b | | 3 | 14.7 | – | 293 | 289 | – | 582 | – | 9.2 | 8.0 | – | 0.83 | Penny 1 |
| 11 | Xinglong | 42 | 186.2 | – | – | 18663 | – | 18663 | – | – | 4.7 | – | 0.06 | Penny 15 |
| 12 | SOAO | 13 | 50.7 | – | 289 | 422 | 522 | 1233 | – | 3.6 | 4.1 | 4.0 | 0.35 | Moffat 25 |
| **Total** | | 351 | 1286.3 | 3075 | 16170 | 43550 | 14648 | 77443 | 5.3 | 6.9 | 5.7 | 5.0 | 0.05 | |

For ORM and Athens Observatory two precision numbers are noted. For ORM, these values refer to the first observation season on the one hand and to the second and third season on the other hand. For Athens Observatory, the two numbers refer to non-binned and binned frames, respectively. $\sigma_{amp,V}=\sigma_V\sqrt{\bar{n}/N_V}$ denotes the noise in the amplitude spectrum of the $V$ light curve, and the PSF function in the table is the analytical function we used to calculate the PSF photometry in DAOPHOT.

and Geneva $UB_1BB_2V_1VG$ measurements of (suspected) $\beta$ Cep stars were collected using the Danish photometer at OAN-SPM and the P7 photometer at ORM respectively. For both photometers, a suitable diaphragm was used to measure the target star only each time. At ORM the photometer was only used from September until October 2006, when technical problems with the Merope CCD were encountered. The precision of the photo-electric photometry reaches 2.5 mmag at OAN-SPM and 10 mmag at ORM.

Figure 1 displays an image with the largest field of view (FOV) of NGC 884 covered by our campaign, denoting the FOVs of all other sites. Since the FOV at ORM was so small, two fields of NGC 884 were observed alternating during the second and third season to cover all the $\beta$ Cep stars known at that time. A world map indicating all the observatories can be found in Saesen et al. (2008). The spread in longitude of the different sites helps to avoid daily alias confusion in the frequency analysis. The effectiveness of this approach is described in Saesen et al. (2009).

The distribution of the data in time per observing site is shown in Fig. 2, while Table 2 contains a summary of the observations. The noted precision indicates the mean standard deviation of the final light curves of ten bright stars with sufficient observations. For ORM, the first number is for the first season and the second number for the second and third season, when another pointing was used. The column with $\sigma_{\mathrm{amp},V}$ denotes the noise in the amplitude spectrum of the $V$ light curve and is calculated as $\sigma_{\mathrm{amp},V}=\sigma_V\sqrt{\pi/N_V}$, with $\sigma_V$ the precision in the $V$ filter and $N_V$ the number of $V$ frames. This is a measure of the relative importance of a certain site to the frequency resolution. It can be deduced that the data from the Białków and Xinglong observatories are the most significant.

The observing strategy was the same for all sites: exposure times were adjusted according to the observing conditions to optimise the signal-to-noise for the known B-type stars, while avoiding saturation and non-linear effects of the camera. The images were preferably taken in focus and with autoguider if present, to keep the star's image sharp to avoid contamination or confusion in the observed half-crowded field. $V$ was the main filter, and where possible, also $B$, $I$ and sometimes $U$ were used. The observations could be taken also during nautical night, as long as the airmass of NGC 884 was below 2.5. A considerable amount of calibration frames was asked for as we were attempting precise measurements. Their description and analysis is discussed in the next section.

## 4. Calibrations

In general, the calibration of the CCD images consisted of bias and dark subtraction, flat fielding and in some cases non-linearity and shutter corrections. They were mostly performed in standard ways and are described in more detail in the sections below.

### 4.1. Bias and dark subtraction

The first step in removing the bias from all images was subtracting the mean level of the overscan region, if available, and trimming this region. This corrected for the average signal introduced by reading the CCD. We tested the relation between this average overscan level and the average bias frame level. Only for one site (ORM) there was a linear dependence which we removed, for all other sites the difference was just a constant offset. After subtracting the overscan level, we corrected for the residual bias pattern, the pixel-to-pixel structure in the read noise on an image. This was mostly done on a nightly basis by taking the mean of several bias frames minus their overscan level (a so-called master bias), which was then subtracted from all other images (darks, flat fields, science frames).

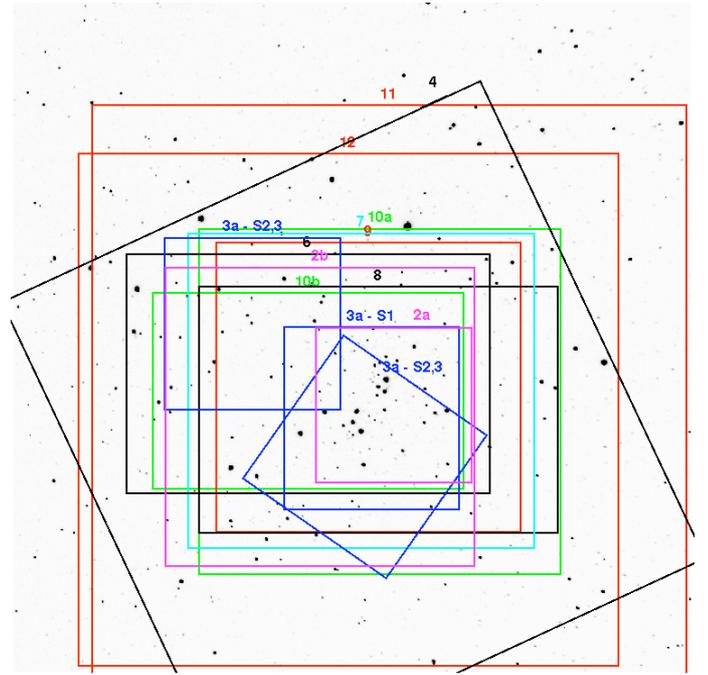

**Fig. 1.** Image of NGC 884 with the largest FOV ($26' \times 26'$ - Michelbach Observatory) in the campaign, with North up and East left. The FOVs of all other observatories are also denoted, where each observatory number refers to the ID of the sites in Table 1.

What remained in the dark frames is the thermal noise, characteristic of the CCD's temperature and exposure time. We constructed a suitable master dark by averaging darks per night and with appropriate duration, and subtracted it to neutralise this effect. The CCDs were temperature controlled so as to keep the dark flux constant, and at several sites its value was even negligible due to the efficient strong cooling of the chip.

### 4.2. Non-linearity correction

The CCD gain can deviate from a perfectly linear reponse, especially for fluxes close to the saturation limit. Since several target stars are bright, we could end up in this non-linear regime. For CCDs where this effect was known, we kept the signal below the level where the non-linearity commences. For unexplored CCDs, we made linearity tests which consisted of flat fields with different exposure times. In most cases, a dome lamp was used with an adjusted intensity so that saturation occurred on a 30-second exposure. A flat-field sequence was then composed of 1, 2, 3, ..., 30 second exposures taken at a random order. In order to reduce uncertainties, at least four such sequences were made to average out the results. Reference images to check the stability of the light source were also collected, but the erratic order of exposure times also helped to mitigate any systematics in the lamp intensity.

As an example we describe here the analysis of the non-linear response of the CCD at OFXB and its correction procedure. The values we use in this description are averaged over the whole image. In the perfect case, the flux rate on the CCD would be constant

$$I_p/t = F,$$

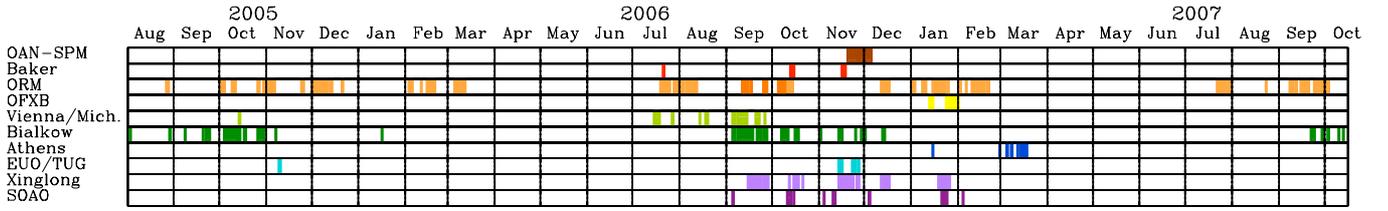

**Fig. 2.** Distribution of the data in time per observing site. The list of observatories from top to bottom goes from west to east.

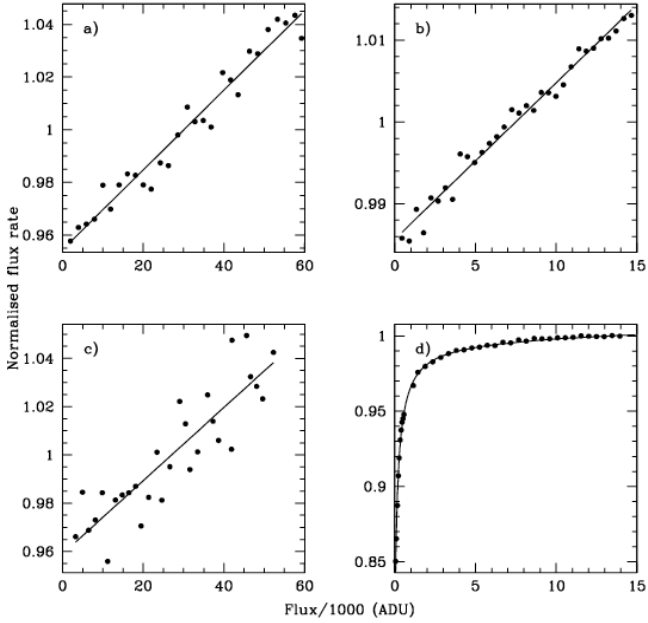

**Fig. 3.** Linearity test for the CCD of a) OFXB b) Michelbach c) SOAO and d) Vienna.

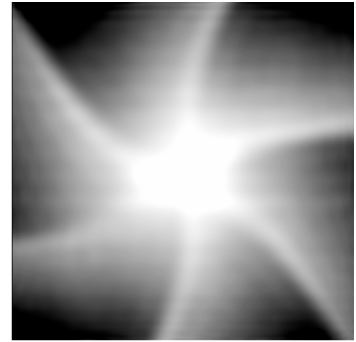

**Fig. 4.** Image showing the calculated $\delta t$ times of the shutter for the CCD at the Mercator telescope at ORM: the middle of the CCD frame is exposed longer than the edges and the shutter blades become visible.

### 4.3. Shutter correction

Since the exposure times at the 1.2-m Mercator telescope (ORM) were short, we corrected the images for the shutter effect. As a mechanical shutter needs time to open and close, some regions in the CCD are exposed longer or shorter depending on their position. To quantify and correct for this unwanted effect, we took a similar series of flat fields as needed for the linearity test, only this time with shorter exposure times. If we divided one very short exposure, where the shutter opening and closing time plays an important role by one of the longer exposures, where the shutter effect is negligible, we could actually nicely see the blades of the shutter (see Fig. 4).

To measure this effect, we handled the problem in the same way as before, only this time the flux rate was not the variable parameter, but the exposure time. In the perfect case, the flux measured in a given pixel would be

$$I_p = Ft,$$

with $F$ again the constant flux rate and $t$ the exposure time. Since at that pixel we have a slightly longer exposure time $\delta t$ due to the mechanical shutter, the measured flux is

$$I_m = F\left(t + \delta t\right).$$

By making a linear least-squares fit between this measured flux and the exposure time of the shutter test sequence, we hence obtained the $\delta t$-value for every pixel. The corrected flux is then given by the subsequent relation:

$$I_p = I_m \frac{t}{t + \delta t}.$$

For all other observing sites except SOAO, the shutter effect was negligible or we could not quantify the effect because of a lack of calibration frames. For SOAO, the correction of the shutter effect was calculated and corrected for in the same way as described above.

with $I_p$ the observed flux, $t$ the exposure time and $F$ a constant. What we measured with the linearity test is a normalised flux rate which increases linearly with the flux (see Fig. 3):

$$\frac{I_m/t}{\langle I_m/t \rangle} = aI_m + b,$$

with $I_m$ the measured flux. We applied a linear least-squares fit to obtain this relation and corrected for it by putting $F$, the desired constant flux rate, equal to the measured mean flux rate: $F_m = \langle I_m/t \rangle$. So obtained the corrected flux by the following relation:

$$I_p = F_m t = \frac{I_m}{aI_m + b}.$$

By further investigation of the linearity tests with corrected fluxes, we noticed a remaining spatial dependent effect that could not be attributed to the shutter (see Sect. 4.3). This was caused by a smooth pixel-to-pixel variation of the coefficients $a$ and $b$ in the fitted linear relation, which we did not account for. Therefore, the linearity correction was performed for every pixel separately, which improved the photometry by a factor of ten.

Three other cameras also reacted in a non-linear way: the ones at Michelbach Observatory and SOAO show the same discrepancy as OFXB, and the CCD at Vienna Observatory exhibits a higher order effect, but only at low fluxes. We corrected for all these CCD effects.

### 4.4. Flat fielding

The final correction applied to the CCD images is flat fielding. It corrects for the pixel-to-pixel sensitivity variations. For every filter and every night, we created a master flat by scaling the flat fields to the normalised level of 1 and taking the mean image while rejecting the extreme values, after checking the flat field's stability by division of one flat by another. Finally the flat field correction was made by dividing the image to be corrected by the master flat. If possible, we took sky flats without tracking (possible stars in the field would then not fall on the same pixels), otherwise dome flats were used. Sometimes we noticed a small difference between dusk and dawn flat fields due to the telescope position. But since we did not dispose of the appropriate calibration frames to correct for it, we took a nightly average. Regions with vignetting on the CCD were trimmed off.

At some observatories, the read-out time of the CCD was very long, so that it was impossible to gather a sufficient number of flat fields in every observed filter per night. In that case, we checked over which time period the flats were stable and combined them together over several nights to construct a master flat. ORM was such an observatory, but the flats from July and August 2006 were not at all stable, caused by the installation of a heat shield in the nitrogen dewar. The flat field change was even noticeable between dusk and dawn and is spatially dependent. Therefore we made a polynomial fit over time for every pixel to interpolate the measured flux variation to gain appropriate flats throughout the night.

## 5. Reduction to differential light curves

The following sections report on the reduction from the calibrated CCD images to differential light curves, which was conducted for each observatory individually (Sect. 5.1-5.4) and the merging of the data of all sites (Sect. 5.5). The reduction method for the photo-electric data taken at ORM with P7 is described in Rufener (1964, 1985). For the reduction of the photo-electric measurements of OAN-SPM, we refer to Poretti & Zerbi (1993) and references therein.

### 5.1. Flux extraction with Daophot

To extract the magnitudes of the different stars on the frames, we used the DAOPHOT II (Stetson 1987) and ALLSTAR (Stetson & Harris 1988) packages.

As a first step we made an extensive master list of 3165 stars in our total field of view by means of a deep and large frame with the best seeing. For this purpose, we utilised the subroutines FIND and PHOT to search for stars and compute their aperture photometry. Although DAOPHOT handles e.g. pixel defects and cosmic rays, the image was checked to reject false star detections. Then we performed profile photometry. The point-spread function (PSF) consisted of an analytical function (appropriate for the instrument, see Table 2) and an empirical table to adapt the PSF in the best way to the image. In general we allowed this function to vary quadratically over the field. Sufficient PSF stars were chosen with PICK for the derivation of the profile shape with the PS subroutine. Subsequently an iterative process was followed to refine this PSF by subtracting neighbouring stars with the calculated profile by ALLSTAR and generating a new PSF with less blended stars until the PSF converged. At last, ALLSTAR fitted all field stars to achieve the best determined CCD coordinates of each star and to obtain their magnitudes. In the end we subtracted all stars in the frame and inspected the resulting image visually. If we could detect any other stars not contained in our master list, we performed the described procedure again on the subtracted image to include them. In addition, a PSF star list was created by selecting isolated and bright stars uniformly spread over the image. The point-spread function calculation was based on these stars instead of further applying the PICK routine.

To convert the star list from one instrument to another, we made a cubic bivariate polynomial transformation of the CCD coordinates X and Y, based on the coordinates of 20 manually cross-identified stars that are well spread over the field. In this way, we obtained a master star list for each CCD. In order to convert an instrument star list of a certain site to each frame, taking only a shift and small rotation into account was sufficient. To do so, we searched for two pre-selected, well-separated, bright but not saturated stars on the image. As a consequence, the strategy to derive the profile photometry as described above could be automated and was used for the fainter stars. This PSF photometry also had the advantage to provide a precise measure of the FWHM (full-width at half-maximum) of the stars, of the mean sky brightness and of the position of the stars in the field.

For the brightest stars, aperture photometry yielded the most precise results. As in our cluster some stars overlap others, we used the routine NEDA to first subtract neighbour stars PSF-wise before determining the aperture photometry. Multiple apertures ranging from 1 until 4 FWHM in steps of 0.25 FWHM and several background radii were tested for every instrument, the one with the most accurate aperture was applied. We tested that using the same aperture for every star yielded best results and so no further correction is needed since the same amount of light is measured for every star.

Some instruments and telescopes dealt with technical problems resulting in distorted images, and sometimes there was bad tracking during the observations. Sporadically, condensation on the cameras occurred and glows were apparent on the images. The weather was not always ideal either. If the images looked really bad, the automatic reduction procedure failed. If the reduction procedure succeeded, but the resulting photometry was inaccurate or inconsistent, the measurements were rejected in the following stage in which we calculated the differential photometry.

### 5.2. Multi-differential photometry

Our aim was to quantify the relative light variations of the measured stars with high precision. In this view, multi-differential photometry, which takes advantage of the ensemble of stars in our field of view, is exactly what we needed. Comparing each star with a set of reference stars indeed reduced the effects of extinction, instrumental drifts and other non-stellar noise sources if their fluctuations in time were consistent over the full frame, but it would still contain the variability information we searched for.

For each instrument we applied an iterative algorithm to compute this relative photometry. In an outer loop, the set of reference stars was iteratively optimised, whereas in an inner loop, the light curve for a given set of reference stars was improved. As a starting point, an initial set of reference stars was extracted from a fixed list of bright stars.

The inner loop of the iteration consisted in adapting the light curve of the reference stars to gain new mean and smaller standard deviation estimates. In the first three iterations, the zero point for image $j$ was calculated as the mean of the variation

of the reference stars

$$\triangle m_j = \frac{\sum_{i=1}^{N} m_{i,j} - \langle m_i \rangle}{N},$$

with $\triangle m_j$ the zero point for frame $j$, $N$ the number of reference stars, $m_{i,j}$ the DAOPHOT magnitude of reference star $i$ in frame $j$ and $\langle m_i \rangle$ the $\sigma$-clipped mean magnitude of this star. Applying this zero point to the reference stars yielded better mean and standard deviation values of their light curve. These values were then used in a following iteration. For the last iteration, a weighted mean was calculated for the zero point

$$\triangle m_j = \sum_{i=1}^{N} \frac{m_{i,j} - \langle m_i \rangle}{\sigma_i^2} / \sum_{i=1}^{N} \frac{1}{\sigma_i^2},$$

where $\sigma_i$ represents the standard deviation of the adapted light curve of reference star $i$ to enhance the contribution of the best stars. We only applied weights in the last iteration to have reliable weights. At different points of the procedure where means were computed, a sigma clipping was used to reject deviating stars.

In the outer loop, the differential magnitudes $m_{c,j}$ of all stars were computed as

$$m_{c,j} = m_{m,j} - \triangle m_j,$$

with $m_{m,j}$ the DAOPHOT magnitude of image $j$ and $\triangle m_j$ the final zero point for this image coming from the inner loop. Based on these adapted magnitudes, new reference stars were selected, i.e., stars with a considerable amount of measurements that have a small standard deviation in their light curve, excluding those that are known to be variable stars. We also checked whether the comparison stars were well distributed over the field.

On average, the final differential photometry was based on 25 reference stars. Images with poor results, where the photometry of the final set of reference stars was used as benchmark, were eliminated at the end. Often this represented 6 to 10% of the total amount of data, but it strongly depended on the instrument.

As a consequence of our choice for bright comparison stars, our approach is optimised for B-type cluster members since they have similar magnitudes and colours as the reference stars. Other stars are typically much fainter, so that their photon noise dominates the uncertainties of the magnitudes.

To improve the overall quality of the data, we removed some outliers with sigma clipping. We used an iterative loop which stops after points were no longer rejected. However, we never discarded several successive measurements as these could originate from an eclipse. Normally the programme converged after a few iterations. The data of the Vienna Observatory ($V$ filter only), Baker Observatory and TUG (T40) were completely rejected since their precision was inferior to the other sites or their photometry was unreliable due to technical problems.

### 5.3. Error determination

As can be noticed in Table 2, we were dealing with a diverse data set concerning the accuracy. To improve the signal-to-noise (S/N) level in a frequency analysis, it was therefore essential to weight the time series (Handler 2003). In order to have appropriate weights, we aimed at getting error estimates per data point which were as realistic as possible. They had to reflect the quality of the measurements in a time series of a certain instrument and at the same time be intercomparable for the different inhomogenous data sets.

In general it is not easy to make a whole picture of the total error budget. The error derivation of DAOPHOT accounts for the readout noise, photon noise, the flat-fielding error and the interpolation error caused by calculating the aperture and PSF photometry. Besides these, there is also scintillation noise, noise caused by the zero point computation by means of the comparison stars and other CCD-related noise sources depending on the colour of the stars, their position etc. Some analytical expressions to evaluate the most important noise sources are available in the literature (e.g., Kjeldsen & Frandsen 1992).

We preferred to work empirically and determined an error approximation together with the relative photometry. We assumed that the standard deviation of the contribution of each reference star to the calculated zero point of a certain image $j$

$$\sigma_j^2 = \frac{\sum_{i=1}^{N}(m_{i,j} - \langle m_i \rangle)^2}{N-1}$$

contained all noise sources. For the bright stars, which have a photon noise comparable to the photon noise of the reference stars, we can use this value directly as the noise estimate of image $j$ since $\sigma_j^2$ already contains a noise component coming from the mean photon noise of all reference stars. The photon noise of faint stars, however, is larger than the one of the relatively bright reference stars, so we have to take the additional photon noise into account in the error estimate. For a given star $i$ and image $j$, our error estimate is

$$\text{noise}_{i,j}^2 = \begin{cases} \sigma_j^2 & \text{if } \epsilon_{i,j} \leq \epsilon_{\text{refstar},j} \\ \sigma_j^2 + \left(\epsilon_{i,j}^2 - \epsilon_{\text{refstar},j}^2\right) & \text{if } \epsilon_{i,j} > \epsilon_{\text{refstar},j}, \end{cases}$$

where $\epsilon$ denotes the photon noise and $\epsilon_{\text{refstar},j}$ was calculated as the mean photon noise of all reference stars used for image $j$. Again, since our comparison stars are bright, these estimates are consistent for other bright stars, and the fainter stars are in any case dominated by photon noise, which is rated by DAOPHOT. Some second order effects like positional depending noises were not accounted for.

We verified if this indeed corresponded to the intercomparable error determination we searched for. To this end, we used the relation between the noise in the time domain $\sigma_{\text{meas}}$, i.e., the mean measured error of the star computed by us, and the noise in the frequency domain $\sigma_{\text{amp}}$, i.e., the amplitude noise in the periodogram, for a given star $i$, given by

$$\langle \text{noise}_{i,j} \rangle = \sigma_{\text{meas},i} = \sqrt{\frac{N}{\pi}} \sigma_{\text{amp},i},$$

with $N$ the number of measurements of star $i$. We checked this correlation for the brightest 300 stars in the field. An example for Białków Observatory can be seen in Fig. 5. It shows that the assumed relation holds. For every other instrument we obtained similar results, except for the Merope camera at the Mercator telescope. Apparently DAOPHOT did not succeed in evaluating the error well, but we could adapt it by simply applying a scaling factor of 3 for the DAOPHOT error for aperture and 1.5 for the PSF photometry. We conclude that we are able to use our error estimates as weights for the data points.

### 5.4. Detrending

Even after taking differential photometry, residual trends can still exist in the light curves, e.g. due to instrumental drifts or changes in the atmospheric transparency. As we were also interested in low frequencies, where the residual trends will cause high noise levels, we wanted to correct the time series for these trends. In order to do so, we applied the Sys-Rem algorithm by Tamuz et al. (2005), which was specifically developped to remove correlated noise by searching for linear systematic effects that are present in the data of a lot of stars. The algorithm minimises the global expression

$$ S^2 = \sum_{ij} \frac{(m_{ij} - \langle m_i \rangle - c_i a_j)^2}{\sigma_{ij}^2}, $$

where $m_{ij}$ and $\sigma_{ij}$ are the magnitude of star $i$ in image $j$ and its error, $\langle m_i \rangle$ is the mean magnitude of star $i$ and $c_i a_j$ is the systematic effect that will be removed, where $c_i$ depends on the star and $a_j$ on the image. These effects can be associated with time, temperature or CCD position.

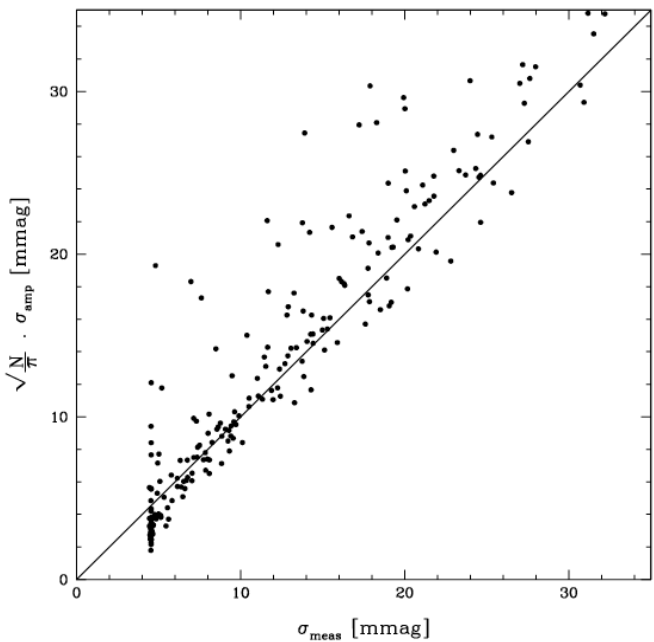

**Fig. 5.** Comparison of the mean measured error $\sigma_{\rm meas}$ with the rescaled amplitude error in the periodogram $\sigma_{\rm amp}$ for the $V$ filter of Białków Observatory. The full line shows the first bisector.

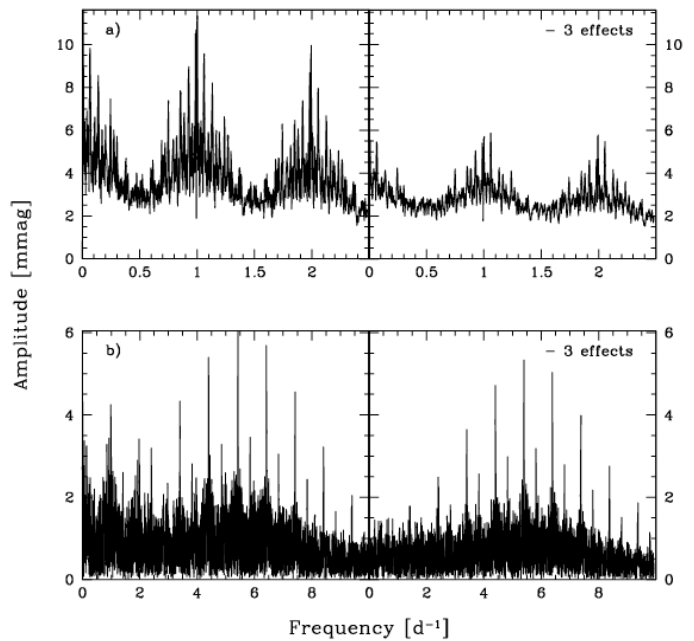

**Fig. 6.** Periodogram examples from Białków Observatory to show the impact of Sys-Rem. a) The average periodogram of the 300 brightest stars in the field, left: without detrending, right: with three effects removed. b) idem for the known $\beta$ Cep star Oo 2246.

Sys-Rem used the error on the measurements to downweigh bad measurements. However, we noticed that including faint stars in the sample to also correct bright stars degrades the photometric results of the latter. Hence we only used the whole set of stars to ameliorate the data of the faintest stars, and we used a sub-sample of bright stars for the application of Sys-Rem to the brightest ones. Stars showing clear long-term variability were also excluded from the sample.

The number of effects to be removed is a free parameter in the algorithm and should be handled with care. Indeed, we wanted to eliminate as many instrumental trends as possible, but certainly no stellar variability. For every instrument we examined the periodograms averaged for the 100 and 300 brightest stars to evaluate when sufficient trends are taken out the data. At the same time the periodograms of the known $\beta$ Cep stars were inspected to make sure the known stellar variability was not affected. An example for Białków Observatory is shown in Fig. 6, and it can be clearly seen that the accuracy improved. In most cases, up to three linear systematic effects were removed.

### 5.5. Merging

At this point the individual time series were fully processed and could be joined per filter ($U$, $B$, $V$, $I$). No phase shifts nor amplitude variations were expected for the data of the various sites for the oscillating B-stars and so only a magnitude shift, which can depend on the star, was computed.

To determine this magnitude shift, we made use of overlapping observation periods. We started with data from Białków Observatory, which holds the most accurate measurements, and added the other instruments one by one, beginning with the site that had the most intervals in common. We smoothed the light curves and interpolated the data points before calculating the mean shift to account for variations in the time series. If no measurements were taken at the same time, the light curves were just shifted by matching their mean values.

## 6. Time-series analysis

To obtain the final light curves, the merged data sets were once more sigma-clipped and detrended with Sys-Rem. The final precision for the brightest stars is 5.7 mmag in $V$, 6.9 mmag in $B$, 5.0 mmag in $I$ and 5.3 mmag in $U$. We also eliminated the light curves of about 25% of the stars, since those stars were poorly measured.

Given the better precision in the $V$ filter and the overwhelming number of data points, the search for variable stars and an automated frequency analysis were carried out in this filter. The $U$-, $B$-, and $I$-filter data will be used in subsequent papers which will present a more detailed frequency analysis of the pulsating stars and eclipsing binaries found in our data.

### 6.1. Detection of variable stars

The following four tools were used to search for variability, whether periodic or not. First, the standard deviation of the light curve gave an impression of the intrinsic variability. However, also the mean magnitude of the star had to be considered, so that the bright variable stars with low amplitudes were selected and not the faint constant stars. To take the brightness of the star into account, we calculated the 'relative standard deviation' of the light curve, i.e., $\sigma_{\rm rel} = (\sigma - \sigma_{\rm av})/\sigma$, where $\sigma_{\rm av}$ is the moving average of the bulk of (constant) stars over their magnitude, as denoted by the grey line in Fig. 7a.

The Abbé test gave another indication of variability. It depends on the point-to-point variations and is sensible to the derivative of the light curve. For a particular star, its value was calculated as

$$\text{Abbé} = \frac{\sum_{i=1}^{N-1}(m_{i+1} - m_i)^2}{2\sum_{i=1}^{N-1}(m_{i+1} - \langle m \rangle)^2},$$

where the sum is taken over the different data points. The value of the Abbé test is close to one for constant stars, larger than one for stars that are variable on shorter time scales than the characteristic sampling of the light curve and significantly smaller than one for variability on longer time scales.

We also determined the reduced $\chi^2$ of the light curves

$$\chi^2_{\rm red} = \frac{\sum_{i=1}^{N}(m_i - \langle m \rangle)^2/e_i^2}{N-1},$$

where $N$ is the number of measurements and $e_i$ is the error on the magnitude $m_i$, as calculated in Sect. 5.3. Its value expresses to what extent the light curve can be considered to have a constant value, the mean magnitude, and so whether there is or not noticeable variability above the noise level. Diagrams of these four indicators ($\sigma$, $\sigma_{\rm rel}$, Abbé and $\chi^2_{\rm red}$) can be found in Fig. 7 and led to the discovery of hundreds of new variables.

A powerful tool to detect periodic variability is frequency analysis. For every star, weighted Lomb-Scargle diagrams (see Sect. 6.2) were calculated from 0 to 50 d$^{-1}$. As this method assumes sinusoidal variations, phase dispersion minimisation diagrams (Stellingwerf 1978) were computed as well. In these periodograms we searched for significant frequency peaks.

In addition, we searched for variable stars by applying the automated classification software by Debosscher et al. (2007). This code was developed primarily for space data (Debosscher et al. 2009), but was previously also applied to the OGLE database of variables (Sarro et al. 2009).

Furthermore, also visual inspection of the light curves ranging over a couple to tens of days gives a strong indication of variability, especially for the brighter stars. After exploring the different diagnostics for variable star detection, we ended up with a list of about 400 stars that were selected for further analysis.

### 6.2. Frequency analysis

As a consequence of the large number of (candidate) variable stars, we applied an automatic procedure for the frequency analysis of the $V$ data, which is a common procedure adopted in the classification of variable star work when treating numerous stars (e.g., Debosscher et al. 2009). This approach cannot be optimal for each individual case, and the results have to be regarded in this way. We thus limit ourselves to report on the dominant frequency only in this work (see Table A.1), even though numerous variables are multi-periodic. An optimal and detailed frequency analysis of the multi-periodic pulsators in the cluster will be the subject of a subsequent paper. For a discussion on the spectral window of the data, we refer to Saesen et al. (2009).

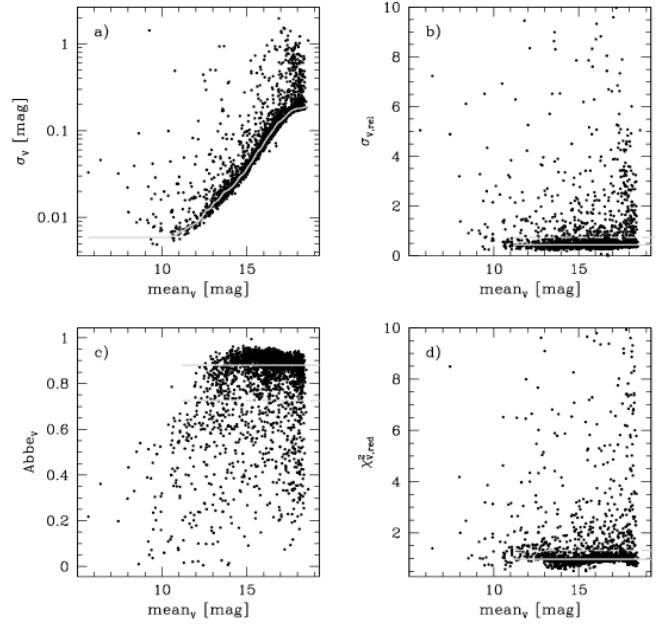

**Fig. 7.** Different diagnostics to detect variability: a) the standard deviation of the light curve, b) the relative standard deviation, c) the Abbé test and d) the reduced $\chi^2$. For the standard deviation, the solid grey line indicates the moving average of the bulk of (constant) stars. For the other indicators, it shows the sigma-clipped average, and the dashed grey line denotes a $3\sigma$ offset to this mean.

Due to changes in observing conditions and instrumentation with different inherent noise levels, there was a large variation in the data quality of the light curve. Following Handler (2003), it was therefore crucial to apply weights when periodograms were computed to suppress bad measurements and enhance the best observing sites. Accordingly, we used the generalised Lomb-Scargle periodogram of Zechmeister & Kürster (2009), where we gave each data point a weight inversely proportional to the square of the error of the measurement: $w_i = 1/e_i^2$. Moreover, this method takes also a floating mean into account besides the harmonic terms in sine and cosine when calculating the Lomb-Scargle fit (Lomb 1976; Scargle 1982). The spectra were computed from 0 d$^{-1}$ to 50 d$^{-1}$ in steps of $1/5T$, where $T$ is the total time span of the light curve.

In these periodograms, a frequency peak was considered as significant when its amplitude was above four times the noise level, i.e., a signal-to-noise ratio greater than 4.0 in amplitude (Breger et al. 1993). The noise at a certain frequency was calculated as the mean amplitude of the subsequent periodogram, i.e., the periodogram of the data prewhitened for the suspected frequency. The interval over which we evaluated the average changed according to the frequency value to account for the increasing noise at lower frequencies: we used an interval of 1 d$^{-1}$ for $f \in [0-3]$ d$^{-1}$, of 1.9 d$^{-1}$ for $f \in [3-6]$ d$^{-1}$, of 3.9 d$^{-1}$ for $f \in [6-11]$ d$^{-1}$ and of 5 d$^{-1}$ for $f \in [11-50]$ d$^{-1}$.

Our strategy for the automatic computations was the following: subsequent frequencies were found by the standard procedure of prewhitening. In up to ten steps of prewhitening, the frequency with the highest amplitude was selected regardless of its S/N, as long as the periodogram contained at least one significant peak. In this way, also residual trends, characterised by low frequencies, and their alias frequencies were removed from the data set, and thus a lower noise level was reached. Starting from the 11th periodogram, peaks with the highest signal-to-noise level were selected. The calculations were stopped once no more significant peaks (S/N > 4) were present in the periodogram. We did not perform an optimisation of the frequency value due to the large calculation time and the excessive number of variables. This can, however, induce the effect of finding almost the same frequency value again after prewhitening.

Since some light curves were not perfectly merged, we adopted the routine described above, excluding frequencies $f = n \pm 0.1\,\mathrm{d}^{-1}$ with $n \in \{0, 1, 2, 3, 4, 5, 6\}\,\mathrm{d}^{-1}$. A constant magnitude shift was calculated to put the residuals of the different observing sites at the same mean level, and this shift was then applied to the original data. Then we repeated this process, now allowing all frequencies. After one more frequency search, the residuals were sigma-clipped and a last frequency analysis was carried out.

The results of this period investigation, which was carried out in the best way we could for an automated analysis, are described from Sect. 8 onward. Frequency values that will be used for stellar modelling need to be derived from a detailed, manual and multi-colour analysis of the 185 variable stars, where the frequency derivation schemes will be optimised according to the pulsator type. This will be presented in subsequent papers for the pulsators.

# 7. Absolute photometry

Absolute photometry of a cluster is a powerful tool to simultaneously determine general cluster parameters like the distance, reddening and age, and stellar parameters for the cluster members like their spectral type and position in the HR diagram. As pointed out in Sect. 2, NGC 884 is a well-studied cluster, but we wanted to take a homogeneous and independent data set of the observed field. As these observations were not directly part of the multi-site campaign, we describe them explicitly together with the specific data reduction and the determination of the cluster parameters. The stellar parameters like cluster membership, effective temperature and gravity, will be deduced and combined with spectroscopy in a subsequent paper.

## 7.1. Observations and data reduction

During seven nights in December 2008 and January 2009, absolute photometry of NGC 884 was taken with the CCD camera Merope at the Mercator telescope at ORM. Since the FoV of this CCD is small, four slightly overlapping fields were mapped to contain as many field stars observed in the multi-site campaign as possible. The seven Geneva filters $U$, $B_1$, $B$, $B_2$, $V_1$, $V$ and $G$ were used in long and short exposures to have enough signal for the fainter stars while not saturating the bright ones. Three to four different measurements spread over different nights were carried out to average out possible variations. The calibration and reduction of these data were performed as described above in Sects. 4 and 5.

In between the measurements of the cluster, Geneva standard stars were observed to account for atmospheric changes. They were carefully chosen to cover different spectral types and observed at various air masses to quantify the extinction of the night. Because the standard stars were very bright and isolated, they were often defocused, and their light was measured by means of aperture photometry with a large radius. A correction factor, determined by bright isolated cluster stars, was applied to the cluster data to quantify their flux in the same way as the standard stars.

First, the measured magnitudes $m_{\mathrm{meas}}$ were corrected for the exposure time $T_{\mathrm{exp}}$ with the formula

$$m = m_{\mathrm{meas}} + 2.5 \log(T_{\mathrm{exp}}).$$

We then determined the extinction in each night for each filter through the measured and catalogue values (Rufener 1988; Burki 2010) of the standard stars

$$m_{\mathrm{meas}} - m_{\mathrm{cat}} = kF_z + \alpha,$$

where $k$ is the extinction coefficient, $F_z$ the airmass of the star and $\alpha$ a constant. We found that the extinction was constant at our precision level during the four hours of observation, and no variations needed to be taken into account. This deduced relation was then applied to the cluster stars.

Despite the effort to match the physical passbands of the CCD system to the reference photometric system, a discrepancy between the observed and catalogue values of the standard stars emerged. It depended on the colours of the stars and was partly due to some flux leakage at red wavelengths of the camera. Therefore an additional transformation to the standard Geneva system is needed. Bratschi (1998) studied the transformation from the natural to the standard Geneva system based on CCD measurements of 242 standard stars. He remarked that the colour-colour residuals clearly show that the connection between these two systems is not a simple relation governed by a single colour index, but rather a relation between the full photometric property of the stars and residuals. Since using all colour indices may not be the most robust and simple way for a transformation as the colour indices are strongly related, he tested whether to use all or a subset of indices. Bratschi (1998) concluded that all indices are needed to remove small local discrepancies of the residuals in relation to the different colours, leading to a significant improvement of the quality of reduction. Following Bratschi (1998), we fitted the most general linear transform to the data of the standard stars

$$C - C_{\mathrm{cat}} = a_U^C.(U - B) + a_V^C.(V - B) + a_{B_1}^C.(B_1 - B) + a_{B_2}^C.(B_2 - B)$$

$$+ a_{V_1}^C.(V_1 - B) + a_G^C.(G - B) + a^C.$$

In this expression, $C$ stands for the six measured Geneva colours on the one hand and the measured $V$ magnitude on the other hand, leading to seven equations in total. The different coefficients $a_i^j$ were determined with a linear least squares fitting and the cluster data were then transformed to the standard system.

In each step described above, we calculated the error propagation. Furthermore, a weighted mean over the different final measurements of each star was taken to average out unwanted variations, leaving us with one final value of the star in each Geneva filter.

## 7.2. Cluster parameters

A number of photometric diagrams in the Geneva system are appropriate to study cluster parameters. These include $(X, Y)$, $(X, V)$, $(B_2 - V_1, B_2 - U)$ and $(B_2 - V_1, V)$, where $X$ and $Y$ are the so-called reddening-free parameters. For their definition and meaning, we refer to Carrier et al. (1999) and references therein. In Carrier et al. (1999), the procedures to deduce the reddening, distance and age of the cluster are explained. We checked the validity of the assumption of uniform reddening in the cluster. For this purpose, we determined the reddening $E(B_2 - V_1)$ of the cluster with the calibration of Cramer (1993) by means of selected B-type cluster members and omitting the known Be stars, since they are additionally reddened by a surrounding disk. The spread in the reddening values was large, having a peak-to-peak difference of $E(B_2 - V_1) = 0.21$, which corresponds to $E(B-V)_J = 0.24$ in the Johnson system following the relations $E(B-V)_J = 0.842\,E(B-V)$ and $E(B_2-V_1) = 0.75\,E(B-V)$ given by Cramer (1984, 1999). Moreover, these reddening variations seemed not randomly spread over the cluster, but show instead spatial correlations. Hence we performed a simulation by randomly permuting the reddening values and measuring the rate of coordinate dependence by the slopes of a linear least-squares fit in function of the CCD X and Y coordinates. This simulation pointed out that the probability of the correlation to be due to noise is smaller than $10^{-5}$ and so proved the non-uniformity of the reddening throughout the cluster, shown in Fig. 8.

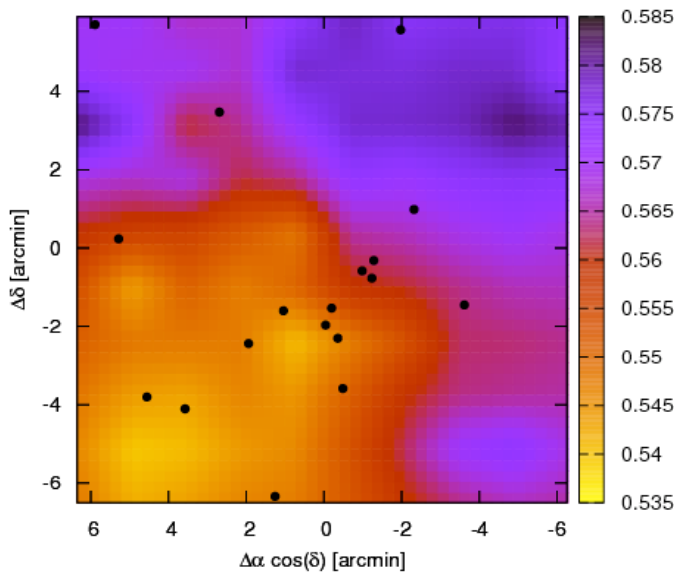

**Fig. 8.** Colour-scale plot of the non-uniform reddening $E(B-V)_J$ in the Johnson system in the cluster field, based on the calibration of Cramer (1993). The centre of the image is $(\alpha, \delta)$=(02h 22m 10s, +57°09′00″). Some bright stars are also denoted.

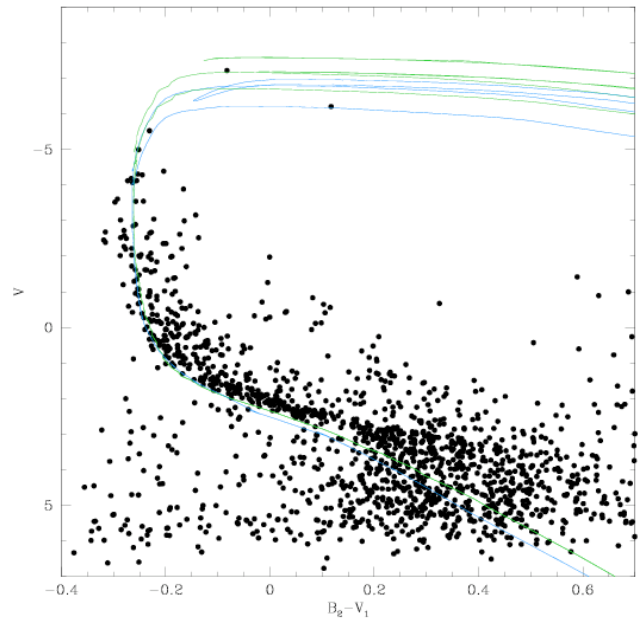

**Fig. 9.** Dereddened and extinction corrected colour-magnitude ($B_2 - V_1$, $V$)-diagram, showing four isochrones: two for log age = 7.1 and 7.2 from Schaller et al. (1992) in blue and two for the same age from Bertelli et al. (1994) in green.

After dereddening all stars by an interpolation of the calibrated reddening values, we adjusted a zero-age main sequence (ZAMS) (Mermilliod 1981) to the photometric diagrams, and this hinted at an over-estimation of the reddening value by $\Delta E(B-V)_J = 0.04 \pm 0.01$ in the Johnson system. The origin of this offset remains unknown. The ZAMS fitting finally resulted in a distance estimate of about 2.2 kpc. To derive the age, we fitted the isochrones of Schaller et al. (1992) and Bertelli et al. (1994). The models of Schaller et al. (1992) we used took standard overshooting, a standard mass loss and solar metallicity into account, and we used the models of Table 5 of Bertelli et al. (1994), suited for solar metallicity. The isochrone fitting lead to a log age of $7.15 \pm 0.15$ where part of the error is due to systematic differences among the theoretical models. Figure 9 denotes the dereddened and extinction corrected colour-magnitude ($B_2 - V_1$, $V$)-diagram, showing a set of isochrones. All the derived values agree well with the literature values mentioned in Sect. 2.

## 8. Variable stars

For the discussion below we divided the variable stars according to their spectral type, determined by the absolute photometry, which was taken for all stars in a homogeneous and consistent way. Thanks to the design of the Geneva small band filters, $B_2 - V_1$ is an excellent effective temperature indicator, and the orthogonal reddening-free parameters $X$ and $Y$ show the spectral type of a star, whether it belongs to the cluster or not (Kunzli et al. 1997). If no absolute photometry was available for the star of interest, ($V - I$, $V$)- and ($B - V$, $V$)- diagrams based on relative photometry were inspected. We also compared literature values for spectral types with the outcome of our photometric diagrams. We classify the stars by their observed variability behaviour: multi-periodic, mono-periodic (possibly with the presence of harmonics) or irregular. The eclipsing binaries are treated in a separate subsection.

All figures for the discovered variable stars can be found in the electronic Appendix A, where we show for each star the $V$ light curve, a phase plot folded with the main frequency, the photometric diagrams, the window function and the generalised Lomb-Scargle diagrams in the different steps of subsequent prewhitening in the $V$-filter. Each separate category of variables will be described here with some typical and atypical examples. For an extensive overview of the properties of the known classes of pulsating stars, we refer to Chapter 2 of Aerts et al. (2009).

Below we adopted a numbering scheme to discuss the stars. Cross referencing to WEBDA numbering (http://www.univie.ac.at/webda/), which is based on the Oosterhoff (1937) numbering and extended to include additional stars coming from other studies, is available in Table A.1 in the electronic Appendix A. This table also contains an overview of the coordinates of the star, its spectral type, its mean Geneva $V$, $B_2 - V_1$ and $B_2 - U$ photometry and its main frequency and amplitude. We recall that we do not list all the frequencies found from the automated analysis in Table A.1. We will provide the final frequencies from a detailed analysis tuned to the various types of pulsators, which may be slightly different from those found here, in a follow-up paper for further use in stellar modelling.

### 8.1. Variable B-type stars

#### 8.1.1. Multi-periodic B-type stars

Figures showing the multi-periodic B-type stars can be found in the electronic Appendix A as Figs A.1–A.72. The classical multi-periodic B-type stars are $\beta$ Cep and SPB stars. $\beta$ Cep stars are early B-type stars showing p-modes with frequency values ranging from 3 to 12 d$^{-1}$. SPB stars are later B-type stars exciting g-modes with lower frequencies from 0.3 to 1 d$^{-1}$. Hybrid B-pulsators, showing at the same time p- and g-modes, also exist.

A typical new $\beta$ Cep star is the early-B star 00011, where we observed three independent frequencies, $f_1 = 4.582$ d$^{-1}$, $f_2 = 5.393$ d$^{-1}$ and $f_5 = 4.449$ d$^{-1}$ and one harmonic frequency, $f_4 = 2f_1$ (see Figs 10–11). We have to be careful with the interpretation of frequency $f_3 = 1.053$ d$^{-1}$: although it deviates more than the resolution (0.001 d$^{-1}$) from 1.003 d$^{-1}$, the application of Sys-Rem can have enhanced this difference.

A typical SPB star is star 02320 (Figs A.63–A.64). This star has three significant frequencies: $f_1 = 0.883$ d$^{-1}$, $f_2 = 0.964$ d$^{-1}$ and $f_3 = 1.284$ d$^{-1}$.

We also found stars exciting p- and g-modes simultaneously. For example star 00030 (Figs A.15–A.16) has two low frequencies $f_1 = 0.340$ d$^{-1}$ and $f_2 = 0.269$ d$^{-1}$, which are typical values for SPB-type pulsations, and it shows one frequency in the $\beta$ Cep star range: $f_4 = 7.189$ d$^{-1}$, while $f_3 = 0.994$ d$^{-1}$ is again too close to 1.003 d$^{-1}$ to be accepted as intrinsic to the star.

We also found some anomalies in the typical classification of variable B-stars. First of all, star 02320 is actually the only SPB candidate in the multi-periodic B-star sample. All other later B-type stars exhibit pulsations with higher frequency values than expected, e.g. the intrinsic frequencies of late-B star 00183 (Figs A.47–A.48) are $f_1 = 3.680$ d$^{-1}$ and $f_2 = 3.904$ d$^{-1}$, which fall in the interval of $\beta$ Cep frequencies. This can however also be a rapidly rotating SPB star, since rotation can induce significant frequency shifts for low-frequency g-modes. However, it could also be a member of a new class of low-amplitude pulsators bridging the red edge of the SPB strip and the blue edge of the $\delta$ Sct strip, as recently found from CoRoT data by Degroote et al. (2009a).

Another peculiar star is star 00024 (Figs A.11–A.12), an early Be-star that revealed two significant frequencies that are low for classical $\beta$ Cep pulsations: $f_1 = 1.569$ d$^{-1}$ and $f_4 = 1.777$ d$^{-1}$. The frequencies are also rather high to be of the SPB class. The phase behaviour does differ from a simple sine wave: looking at other known Be stars, this behaviour seems quite typical for these stars, since frequencies and/or amplitudes may change over time as recently found from uninterrupted CoRoT photometry (e.g., Huat et al. 2009; Neiner et al. 2009).

Star 02451 (Figs A.69–A.70) showed clear variations around 25 and 29 d$^{-1}$, frequency values we would expect for $\delta$ Sct stars. Because of its B-star appearance in the photometric diagrams, the star is therefore probably no cluster member.

#### 8.1.2. Mono-periodic B-type stars

The figures of the frequency analysis for mono-periodic B-type stars can be found in the electronic Appendix A in Figs A.73–A.150. For the mono-periodic cases, the same nature of pulsations as described above is also observed. However, because we did not find more than one independent frequency, we cannot be sure that the variations are caused by oscillations. Possibilities like spotted stars or ellipsoidal binaries cannot be ruled out, especially if we also deduce harmonics.

Star 00082 (Figs A.95–A.96) is a typical example: $f_1 = 0.639$ d$^{-1}$ and $f_2 = 2f_1$ and the phase plot folded with the main frequency is therefore clearly not sinusoidal. Because of the high amplitude of $f_1$, it is possible that we are dealing with non-sinusoidal SPB-oscillations, but it could also be a spot.

In this sample we also have some B-type supergiants that vary with SPB-type periods. These were originally discovered by Waelkens et al. (1998) in the Hipparcos mission and were studied in detail by Lefever et al. (2007). For instance, star 00008 (Figs A.75–A.76), is listed as B2I supergiant (Slesnick et al. 2002) and has one clear frequency at $f_2 = 0.211$ d$^{-1}$. Possibly $f_3 \approx 2f_2$ is also present, which is not so surprising since the oscillations of supergiant stars are often non-sinusoidal.

#### 8.1.3. Irregular B-type stars

For some stars, the automatic frequency analysis failed, since long-term trends dominate the light curves. This is the case for Be stars that undergo an outburst. For instance, the light curve of star 00009 (Fig. 12) revealed a huge outburst in the beginning of the second observing season and a smaller one in the third observing season. Smaller variations are possibly still hidden in the time series, but a more detailed analysis should point this out.

Other Be stars without significant frequencies do not always show an outburst, but frequency and/or amplitude changes prevented us from detecting the frequency values. Zooms of the light curve sometimes show clear periodicity on time-scales of several hours. Their light curves and some zooms are shown in Figs A.152–A.159. Star 00003 behaves in the same way (see Fig. A.151), but is not known to be a Be star.

### 8.2. A- and F-type stars

#### 8.2.1. Multi-periodic A- and F-type stars

Figures showing the multi-periodic A- and F-type stars can be found in Figs A.160–A.197 of the electronic Appendix A. These stars are probably $\delta$ Sct stars, exciting p-modes with frequencies ranging from 4 d$^{-1}$ to 50 d$^{-1}$ or $\gamma$ Dor stars with g-modes in the frequency interval from 0.2 d$^{-1}$ to 3 d$^{-1}$.

For example, A-type star 00315 (Figs A.168–A.169) shows two significant $\delta$ Sct frequencies $f_1 = 35.337$ d$^{-1}$ and $f_4 = 21.123$ d$^{-1}$. F-type star 00371 (Figs A.180–A.181) shows four distinct $\gamma$ Dor frequencies between 0.243 d$^{-1}$ and 1.083 d$^{-1}$.

As for the B-type stars, hybrid pulsators showing both p-and g-modes were also discovered among the A-F type stars. F-type star 00187 (Figs A.160–A.161) could be such a star. After prewhitening for $f_1 = 1.888$ d$^{-1}$ and $f_2 = 0.983$ d$^{-1}$, the higher frequencies $f_5 = 8.135$ d$^{-1}$ and $f_{11} = 42.565$ d$^{-1}$ appeared.

An interesting case is A-star 00344 (Figs A.174–A.175). In this star we observed the frequencies $f_1 = 0.840$ d$^{-1}$ and $f_2 = 0.875$ d$^{-1}$, which are too low for a $\delta$ Sct star. These periodicities could be of SPB or $\gamma$ Dor nature, but following the absolute photometry, the star's spectral type is incompatible with this. The same holds for the star 00577 (Figs A.230-A.231). Here the photometric diagrams indicated that this is an F-star, but the periodogram shows a higher frequency value than for the $\gamma$ Dor pulsators: $f_3 = 8.258$ d$^{-1}$.

### 8.2.2. Mono-periodic A- and F-type stars

Mono-periodic $\delta$ Sct and $\gamma$ Dor stars also exist. The $\delta$ Sct stars can be unambiguously separated from rotating variables on the basis of their short periods, while $\gamma$ Dor stars cannot, due to their longer periods. All examples can be found in Figs A.198-A.245. We remark that no harmonics were found for the mono-periodic A-type stars.

### 8.3. Other variable stars

Besides stars with another spectral type than B, A or F, stars for which the spectral type is uncertain or where no information on this is at our disposal are also treated in this subsection. Their frequency analyses are summarised in Figs A.246-A.285 for the multi-periodic stars and in Figs A.286-A.325 for the mono-periodic stars. Using the frequency values alone, it seems not possible to reliably classify the stars.

For the main-sequence G-type stars we only expected solar-like oscillations, spots, or ellipsoidal binaries, but we had some cases with a few close frequency values and sometimes nearly a harmonic frequency. For example star 00681 (Figs A.252-A.253) has the significant frequencies $f_1 = 1.818$ d$^{-1}$ and $f_2 = 1.808$ d$^{-1}$. This is probably a sign of bad prewhitening since we did not optimise the frequency value before prewhitening.

In addition, six irregular variable stars were also found. They are listed as supergiants in SIMBAD: 00001 (M4I), 00002 (B2I), 02427 (M1I), 02428 (M1I) and 03165 (A2I), except 03164 (sdM1). However, the spectral type of star 03164 is probably wrong in SIMBAD, and this star is listed as M0I supergiant in Slesnick et al. (2002). For these stars we avoid to show the photometric diagrams, since the values are uncertain. For completeness, we do show the light curves and some zooms in Figs A.326-A.331.

### 8.4. Eclipsing binaries

Apart from the two known cases announced in Krzesiński & Pigulski (1997), we discovered six new and four candidate eclipsing binaries. Classical frequency analysis methods did not succeed to find their periods, so we estimated them from the eclipse minima and by visual inspection of the phase plots. The star numbers of the confirmed eclipsing binaries together with their periods are given in Table 3. Their phase plots can be found in the electronic Appendix A in Figs A.332-A.339. The four candidate eclipsing binaries, stars 00033, 00045, 00110 and 00316, show dips in their light curves that could point to an eclipse. Extracts of these parts of their light curve are shown in Figs A.340-A.343. A separate paper including long-term spectroscopy is in preparation and is devoted to star 00010 (Southworth et al., in preparation), while the other eclipsing binaries we discovered will be modelled together in a subsequent paper. The studies on these eclipsing binaries will be performed to deduce more cluster characteristics, which will serve as input for any future modelling.

**Table 3.** The (candidate) eclipsing binaries.

| Star ID | WEBDA | Period (d) | Remark |
|---------|-------|-----------:|--------|
| 00010 | 2311 | 25.53 | known |
| 00028 | 2433 | 0.95 | new |
| 00033 | 2232 | / | new |
| 00045 | 2191 | / | new |
| 00048 | 2251 | 11.53 | new |
| 00051 | 2351 | 11.61 | new |
| 00058 | 2301 | 1.55 | known |
| 00110 | 2275 | / | new |
| 00316 | 2490 | / | new |
| 02217 | 1924 | 15.23 or 30.46 or 60.92 | new |
| 02227 | 1923 | 8.84 | new |
| 02302 | 2646 | 8.20 | new |

## 9. Summary and conclusion

We have carried out an extensive multi-site campaign to gather time-resolved multi-colour CCD photometry of a field of the open cluster NGC 884. The aim was to search for variable stars, in particular B-type pulsators. In total, an international team consisting of 61 observers used 15 different instruments attached to 13 telescopes to collect almost 77 500 CCD images and 92 hours of photo-electric data in $U$, $B$, V and $I$ filters.

We performed the calibration of the CCD images by standard bias and dark subtractions, non-linearity and shutter corrections and flat fielding. We extracted the fluxes of the stars with the DAOPHOT II (Stetson 1987) and ALLSTAR (Stetson & Harris 1988) packages, in which we combined PSF and aperture photometry. We applied multi-differential photometry to correct for atmospheric extinction and detrended the data sets with the Sys-Rem algorithm by Tamuz et al. (2005). We succeeded in deducing realistic and inter-comparable error estimates for the measurements. For the brightest stars, we obtained an overall $V$ accuracy of 5.7 mmag, 6.9 mmag in $B$, 5.0 mmag in $I$ and 5.3 mmag in $U$.

The search for variable stars among the 3165 observed stars was based on several indicators like the standard deviation of the light curve, the Abbé test, the reduced $\chi^2$ value, frequency analysis and visual inspection. About 400 candidate variable stars were subjected to an automated weighted frequency analysis revealing 36 multi-periodic and 39 mono-periodic B-stars, 19 multi-periodic and 24 mono-periodic A- and F-stars and 20 multi-periodic and 20 mono-periodic variable stars of unknown nature. Moreover, 15 irregular variable stars were found, of which eight are Be stars and six (super)giant stars. We also detected, apart from two known cases, six new and four candidate eclipsing binaries.

The interpretation of these variable stars is not always straightforward and needs further investigation. In general, it seems that we cannot rely anymore on the simple classification of B-type stars by means of the theoretical instability domains for $\beta$ Cep and SPB stars as in e.g. Miglio et al. (2007). This was also pointed out by Degroote et al. (2009a) based on their analysis of B-type stars in the CoRoT exoplanet field data. We arrive at the same conclusions, derived from an independent data set with totally different characteristics: numerous stars have frequency ranges extending from the gravity-mode range corresponding to periods of days to the pressure-mode regime corresponding to periods of a few hours. A part of the anomalies in the classical categorisation can be explained by rotation, since, as discussed in Sect. 2, it is well known that the average rotational velocities for the brightest stars in this cluster is high (Slettebak 1968) and this affects the observed frequency values. However, the com-plication probably also arises because both the CoRoT data and our cluster data exceed the previous data sets of B-type pulsators by far in terms of number of targets and photometric precision in the amplitude spectra. We have thus lowered the threshold of finding new low-amplitude pulsators, and these seem to come in more flavours than anticipated so far.

Earlier variability studies on NGC 884 showed the detection of only very few B-type pulsators: there are two confirmed and two candidate $\beta$ Cep star s in the cluster (see Sect. 2). Waelkens et al. (1990) and Pigulski et al. (2001) were puzzled by this fact, since NGC 884 has roughly the same age as NGC 3293, the southern open cluster which yields most known $\beta$ Cep stars (Stankov & Handler 2005). Waelkens et al. (1990) proposed that these observations lend support to the old idea that large rotational velocities tend to be incompatible with the $\beta$ Cep phenomenon. Pigulski et al. (2001) speculated that the metallicity gradient in the Galaxy may be responsible for this difference. Our data set revealed 36 multi-periodic and 39 mono-periodic B-type variables, so the lack of detection of these oscillators in the past was maybe an observational constraint due to limited precision and field-of-view, as well as to short time bases and too low duty cycles, certainly when noting that most amplitudes of the variations we detect are low.

We identified periodic changes in several Be stars. We also noticed that their phase plot is quite chaotic: it shows more scatter around the mean light curve in comparison with other pulsating stars with the same brightness (e.g., $\beta$ Cep stars). It could originate from frequencies and/or amplitudes that change over time. Some other periodically varying B-type stars have the same photometric behaviour, suggesting a similar Be nature, although we do not have direct spectroscopic evidence. This erratic behaviour in the phase plot was already pointed out by Jerzykiewicz et al. (2003). These findings on Be pulsators again fully agree with recent results from the CoRoT mission, where outbursts could be interpreted as beating between pressure- and gravity-modes with time dependent amplitudes (e.g., Huat et al. 2009; Neiner et al. 2009).

In a subsequent paper we will perform a manual and more detailed frequency analysis of the pulsators, especially the B-type stars. For the moment we did not find clear connections between the dominant frequency and various observed properties of the stars, but we will come back on this issue once we studied the different classes of variable stars in detail. After a detailed frequency analysis we can also take all detected frequencies into account, instead of only the dominant one which is treated in this paper. A mode identification will also be carried out based on the multi-colour photometry, to determine the degree $\ell$ of the oscillations. The eclipsing binaries will be treated as well in more detail in another paper to deduce more cluster characteristics. Hereafter, an asteroseismic modelling of the pulsators in the cluster will be performed, assuming that they have the same age and that they had the same chemical composition at birth since they were born out of the same cloud. Given the numerous B-type pulsators discovered in this cluster, an in-depth evaluation of the stellar evolution models seems very promising.

*Acknowledgements.* The research leading to these results has received funding from the European Research Council under the European Community's Seventh Framework Programme (FP7/2007–2013)/ERC grant agreement n°227224 (PROSPERITY), from the Research Council of K.U.Leuven grant agreement GOA/2008/04, from the Fund for Scientific Research of Flanders grant G.0332.06 and from the European Helio- and Asteroseismology Network (HELAS), a major international collaboration funded by the European Commission's Sixth Framework Programme. AP, GM and ZK were supported by the NN203 302635 grant from the MNiSW. KU acknowledges financial support from a *European Community Marie Curie Intra-European Fellowship*, contract number MEIF-CT-2006-024476. This research has made use of the WEBDA database, operated at the Institute for Astronomy of the University of Vienna, as well as NASA's Astrophysics Data System and the SIMBAD database, operated at CDS, Strasbourg, France.

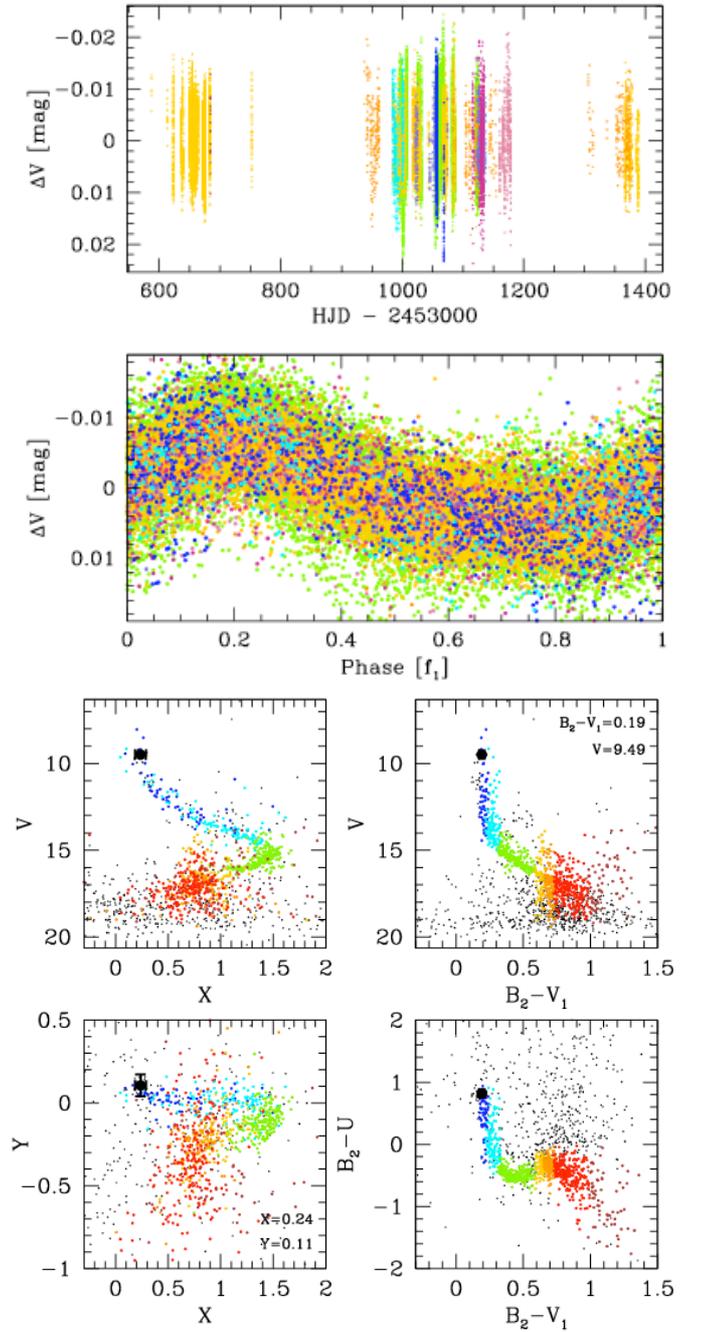

**Fig. 10.** Light curve (top), phase plot (middle) and photometric diagrams (bottom) of star 00011. The light curve and phase plot are made from the *V*-filter observations and the different colours denote the different observing sites: ORM – orange, OFXB – dark pink, Michelbach – light blue, Białków – yellow, Athens – light pink, EUO – dark blue, TUG – brown, Xinglong – green and SOAO – purple. The phase plot is folded with the main frequency, denoted in the X-label and whose value is listed in Table A.1. The different colours in the photometric diagrams indicate the spectral types: B0-B2.5 – dark blue, B2.5-B9 – light blue, A – green, F0-F2 – yellow, F3-F5 – orange, F6-G – red, K-M – brown. The big dot with error bars shows the position of star 00011 in these figures.

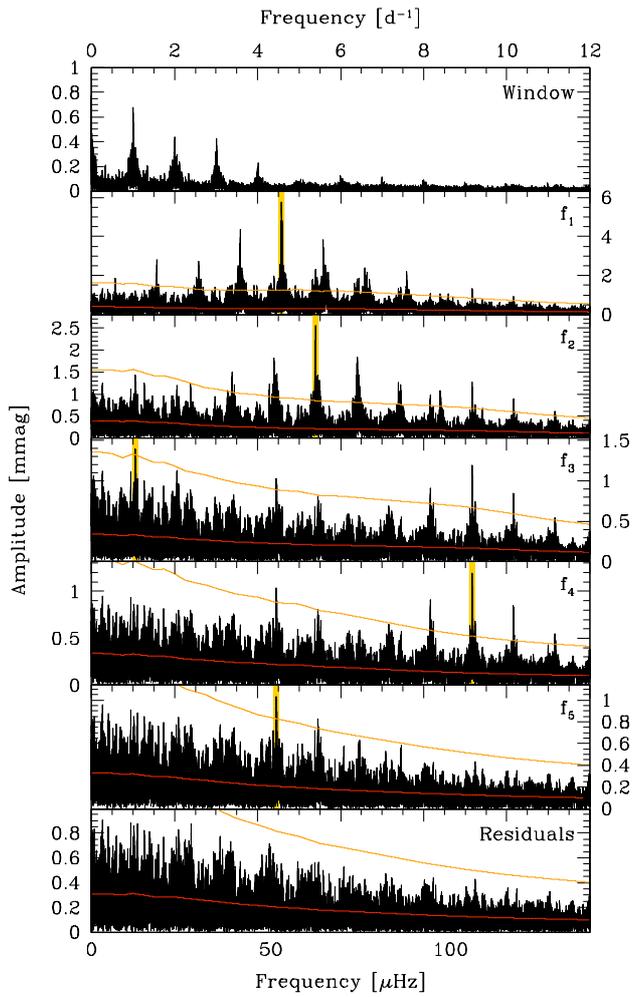

**Fig. 11.** Frequency analysis for star 00011. We show the window function (top) and the generalised Lomb-Scargle periodograms in the different steps of subsequent prewhitening in the *V*-filter. The detected frequencies are marked by a yellow band, the red line corresponds to the noise level and the orange line to the S/N=4 level.

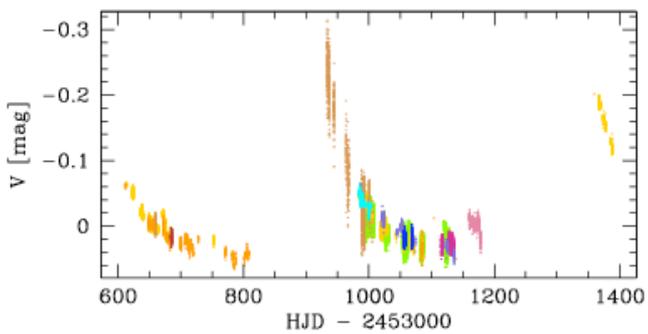

**Fig. 12.** Light curve of Be-star 00009, showing outbursts. The different colours denote the different observing sites as in Fig. 10, the light brown data come from the Vienna Observatory.

**Appendix A: Additional tables and figures**

**Table A.1.** Overview of the variable stars treated in Sects. 8.1-8.3. The subsequent columns denote our star identification used in this paper, the WEBDA star number if available, the $\alpha$ and $\delta$ coordinates of the star, the spectral type taken from SIMBAD, the mean Geneva $V$, $B_2 - V_1$ and $B_2 - U$ photometry, the main frequency and its amplitude and the section and figures of this paper where the star is treated. Sections 8.1.1, 8.1.2 and 8.1.3 refer to the multi-periodic, mono-periodic and irregular B stars, Sects. 8.2.1 and 8.2.2 to the multi- and mono-periodic A and F stars and Sect. 8.3 to the other variable stars.

| Star ID | WEBDA | $\alpha$ (h m s) | $\delta$ (° ′ ″) | SpT | $V$ | $B_2 - V_1$ | $B_2 - U$ | Frequency (d⁻¹) | (μHz) | Ampl. (mmag) | Section | Figures |
|---|---|---|---|---|---|---|---|---|---|---|---|---|
| 00004 | 2299 | 02 22 08.55 | +57 07 28 | B0.5IV | 9.15 | 0.24 | 0.91 | 3.145 | 36.40 | 6.3 | 8.1.1 | A.1-A.2 |
| 00011 | 2444 | 02 22 29.85 | +57 12 28 | | 9.49 | 0.19 | 0.82 | 4.582 | 53.03 | 5.8 | 8.1.1 | A.3-A.4 |
| 00012 | 2246 | 02 22 02.76 | +57 08 25 | B2III | 9.94 | 0.16 | 0.81 | 5.429 | 62.84 | 5.6 | 8.1.1 | A.5-A.6 |
| 00013 | 2572 | 02 22 48.95 | +57 09 14 | | 9.99 | 0.20 | 0.76 | 4.415 | 51.10 | 5.2 | 8.1.1 | A.7-A.8 |
| 00014 | 2488 | 02 22 36.28 | +57 04 54 | B1V | 9.96 | 0.16 | 0.91 | 6.168 | 71.39 | 6.9 | 8.1.1 | A.9-A.10 |
| 00024 | 2242 | 02 22 02.45 | +57 09 20 | B1Vnp | 10.93 | 0.23 | 0.65 | 1.569 | 18.16 | 12.8 | 8.1.1 | A.11-A.12 |
| 00027 | 2185 | 02 21 55.55 | +57 05 58 | B3V | 10.95 | 0.17 | 0.52 | 0.625 | 7.23 | 1.2 | 8.1.1 | A.13-A.14 |
| 00030 | 2114 | 02 21 46.71 | +57 07 27 | B2V | 11.05 | 0.18 | 0.72 | 0.340 | 3.94 | 7.6 | 8.1.1 | A.15-A.16 |
| 00041 | 2091 | 02 21 44.46 | +57 10 52 | B3 | 11.62 | 0.28 | 0.48 | 2.511 | 29.06 | 2.6 | 8.1.1 | A.17-A.18 |
| 00045 | 2191 | 02 21 56.93 | +57 12 20 | | 11.65 | 0.28 | 0.51 | 0.328 | 3.80 | 4.0 | 8.1.1 | A.19-A.20 |
| 00055 | 2094 | 02 21 44.38 | +57 08 54 | B2 | 11.85 | 0.19 | 0.69 | 0.376 | 4.35 | 4.1 | 8.1.1 | A.21-A.22 |
| 00073 | 2507 | 02 22 38.85 | +57 11 44 | | 12.49 | 0.20 | 0.47 | 2.407 | 27.86 | 2.8 | 8.1.1 | A.23-A.24 |
| 00079 | 2245 | 02 22 02.77 | +57 08 52 | B5 | 12.49 | 0.20 | 0.50 | 2.670 | 30.91 | 2.3 | 8.1.1 | A.25-A.26 |
| 00084 | 2110 | 02 21 45.95 | +57 05 01 | | 12.65 | 0.27 | 0.33 | 2.600 | 30.10 | 1.7 | 8.1.1 | A.27-A.28 |
| 00088 | 2253 | 02 22 03.48 | +57 07 08 | B7 | 12.66 | 0.23 | 0.35 | 0.714 | 8.27 | 9.1 | 8.1.1 | A.29-A.30 |
| 00100 | 2349 | 02 22 14.63 | +57 06 13 | B7 | 12.83 | 0.18 | 0.41 | 2.699 | 31.24 | 2.3 | 8.1.1 | A.31-A.32 |
| 00103 | 2429 | 02 22 27.25 | +57 09 07 | | 12.93 | 0.17 | 0.44 | 1.083 | 12.53 | 3.5 | 8.1.1 | A.33-A.34 |
| 00115 | 2319 | 02 22 10.62 | +57 06 40 | B8 | 13.11 | 0.19 | 0.31 | 3.065 | 35.47 | 5.6 | 8.1.1 | A.35-A.36 |
| 00118 | 2345 | 02 22 13.69 | +57 05 17 | B3V | 13.08 | 0.18 | 0.43 | 2.000 | 23.15 | 9.2 | 8.1.1 | A.37-A.38 |
| 00128 | 2267 | 02 22 04.90 | +57 06 01 | B7 | 13.25 | 0.20 | 0.29 | 2.603 | 30.13 | 3.6 | 8.1.1 | A.39-A.40 |
| 00138 | 2228 | 02 22 00.35 | +57 04 36 | | 13.57 | 0.32 | -0.18 | 3.799 | 43.97 | 1.1 | 8.1.1 | A.41-A.42 |
| 00145 | 2515 | 02 22 39.65 | +57 04 49 | | 13.48 | 0.20 | 0.32 | 2.532 | 29.31 | 5.8 | 8.1.1 | A.43-A.44 |
| 00166 | 2406 | 02 22 22.42 | +57 11 02 | | 13.85 | 0.27 | 0.08 | 3.707 | 42.90 | 1.9 | 8.1.1 | A.45-A.46 |
| 00183 | 2324 | 02 22 11.26 | +57 06 33 | A0 | 13.91 | 0.22 | 0.03 | 3.680 | 42.59 | 4.0 | 8.1.1 | A.47-A.48 |
| 00196 | 2531 | 02 22 42.97 | +57 12 19 | | 14.10 | 0.26 | -0.07 | 3.664 | 42.41 | 3.7 | 8.1.1 | A.49-A.50 |
| 00202 | 2410 | 02 22 23.22 | +57 10 05 | A0 | 14.13 | 0.25 | -0.04 | 3.342 | 38.68 | 3.2 | 8.1.1 | A.51-A.52 |
| 00218 | 2323 | 02 22 11.12 | +57 06 24 | A0 | 14.24 | 0.26 | -0.09 | 3.974 | 46.00 | 2.2 | 8.1.1 | A.53-A.54 |
| 02230 | 2633 | 02 23 01.00 | +57 07 17 | | | | | 8.236 | 95.32 | 2.4 | 8.1.1 | A.55-A.56 |
| 02293 | 2448 | 02 22 29.02 | +57 01 43 | | | | | 0.497 | 5.75 | 7.4 | 8.1.1 | A.57-A.58 |
| 02299 | 2649 | 02 23 04.18 | +57 07 39 | Be | | | | 2.244 | 25.97 | 14.8 | 8.1.1 | A.59-A.60 |
| 02300 | 2601 | 02 22 56.81 | +57 11 44 | | 10.64 | 0.30 | 0.59 | 6.702 | 77.57 | 2.4 | 8.1.1 | A.61-A.62 |
| 02320 | 2089 | 02 21 43.00 | +57 00 54 | | | | | 0.883 | 10.22 | 5.8 | 8.1.1 | A.63-A.64 |
| 02430 | 2694 | 02 23 11.62 | +57 10 20 | | | | | 3.733 | 43.20 | 5.4 | 8.1.1 | A.65-A.66 |
| 02436 | 2520 | 02 22 41.81 | +57 16 59 | | | | | 5.657 | 65.48 | 2.7 | 8.1.1 | A.67-A.68 |
| 02451 | 2086 | 02 21 44.00 | +57 18 21 | | | | | 25.275 | 292.54 | 2.5 | 8.1.1 | A.69-A.70 |
| 02540 | 2753 | 02 23 22.00 | +57 00 26 | | | | | 2.233 | 25.85 | 7.6 | 8.1.1 | A.71-A.72 |
| 00006 | 2371 | 02 22 17.67 | +57 07 24 | B2III | 9.20 | 0.19 | 0.81 | 0.384 | 4.44 | 20.2 | 8.1.2 | A.73-A.74 |
| 00008 | 2235 | 02 22 00.92 | +57 08 14 | B1V | 9.37 | 0.19 | 0.80 | 0.211 | 2.44 | 1.0 | 8.1.2 | A.75-A.76 |
| 00017 | 2566 | 02 22 48.10 | +57 12 01 | B1ne | 10.47 | 0.32 | 0.74 | 2.552 | 29.54 | 13.3 | 8.1.2 | A.77-A.78 |
| 00018 | 2262 | 02 22 04.55 | +57 10 38 | B2Vn | 10.55 | 0.24 | 0.77 | 3.072 | 35.56 | 5.7 | 8.1.2 | A.79-A.80 |
| 00029 | 2377 | 02 22 19.20 | +57 11 36 | B | 11.05 | 0.22 | 0.74 | 1.388 | 16.06 | 1.6 | 8.1.2 | A.81-A.82 |
| 00039 | 2139 | 02 21 49.64 | +57 08 21 | B2V | 11.36 | 0.18 | 0.76 | 1.477 | 17.10 | 1.7 | 8.1.2 | A.83-A.84 |
| 00042 | 2462 | 02 22 32.76 | +57 08 41 | | 11.49 | 0.18 | 0.65 | 1.251 | 14.48 | 2.1 | 8.1.2 | A.85-A.86 |
| 00043 | 2372 | 02 22 18.56 | +57 11 50 | | 11.60 | 0.22 | 0.67 | 0.766 | 8.86 | 3.3 | 8.1.2 | A.87-A.88 |
| 00057 | 2579 | 02 22 50.28 | +57 08 50 | B3 | 11.91 | 0.19 | 0.56 | 2.282 | 26.41 | 11.0 | 8.1.2 | A.89-A.90 |
| 00068 | 2189 | 02 21 56.70 | +57 12 49 | | 12.29 | 0.27 | 0.47 | 0.104 | 1.20 | 3.2 | 8.1.2 | A.91-A.92 |
| 00072 | 2352 | 02 22 15.16 | +57 09 40 | B6 | 12.32 | 0.16 | 0.41 | 0.291 | 3.37 | 1.8 | 8.1.2 | A.93-A.94 |
| 00082 | 2455 | 02 22 30.35 | +57 03 51 | | 12.54 | 0.18 | 0.51 | 0.639 | 7.40 | 15.1 | 8.1.2 | A.95-A.96 |
| 00086 | 2285 | 02 22 07.23 | +57 12 11 | | 12.64 | 0.24 | 0.56 | 0.623 | 7.21 | 9.6 | 8.1.2 | A.97-A.98 |
| 00095 | 2006 | 02 21 28.41 | +57 08 34 | | 12.81 | 0.24 | 0.35 | 2.703 | 31.29 | 2.6 | 8.1.2 | A.99-A.100 |
| 00099 | 2309 | 02 22 09.57 | +57 06 33 | B7 | 12.81 | 0.19 | 0.38 | 5.741 | 66.44 | 2.1 | 8.1.2 | A.101-A.102 |
| 00104 | 2037 | 02 21 33.75 | +57 03 56 | | 13.09 | 0.27 | 0.21 | 0.164 | 1.90 | 16.0 | 8.1.2 | A.103-A.104 |
| 00123 | 1980 | 02 21 24.83 | +57 02 25 | | | | | 6.266 | 72.52 | 3.7 | 8.1.2 | A.105-A.106 |
| 00131 | 2350 | 02 22 14.89 | +57 09 38 | B7 | 13.36 | 0.22 | 0.31 | 2.632 | 30.46 | 5.1 | 8.1.2 | A.107-A.108 |
| 00139 | 1990 | 02 21 25.71 | +57 04 22 | | 13.52 | 0.29 | 0.12 | 2.974 | 34.42 | 5.2 | 8.1.2 | A.109-A.110 |
| 00142 | 2116 | 02 21 46.91 | +57 09 54 | B8 | 13.50 | 0.29 | 0.17 | 3.213 | 37.19 | 1.7 | 8.1.2 | A.111-A.112 |
| 00146 | 2426 | 02 22 25.98 | +57 06 53 | B8V | 13.75 | 0.25 | 0.07 | 0.752 | 8.70 | 1.5 | 8.1.2 | A.113-A.114 |
| 00162 | 2414 | 02 22 23.84 | +57 08 26 | A2 | 13.83 | 0.22 | -0.27 | 0.437 | 5.06 | 7.7 | 8.1.2 | A.115-A.116 |
| 00163 | 2524 | 02 22 41.66 | +57 10 28 | | 13.86 | 0.30 | -0.02 | 3.222 | 37.29 | 2.0 | 8.1.2 | A.117-A.118 |
| 00188 | 2562 | 02 22 47.45 | +57 11 27 | | 14.20 | 0.42 | -0.14 | 2.574 | 29.79 | 7.7 | 8.1.2 | A.119-A.120 |
| 00201 | 2370 | 02 22 17.74 | +57 06 13 | B9V | 14.02 | 0.21 | 0.03 | 3.274 | 37.90 | 4.0 | 8.1.2 | A.121-A.122 |
| 00238 | 2482 | 02 22 35.99 | +57 12 32 | | 14.44 | 0.19 | -0.20 | 0.702 | 8.12 | 2.8 | 8.1.2 | A.123-A.124 |

**Table A.1.** Continued.

| Star ID | WEBDA | $\alpha$ (h m s) | $\delta$ (° ′ ″) | SpT | $V$ | $B_2 - V_1$ | $B_2 - U$ | Frequency (d$^{-1}$) | ($\mu$Hz) | Ampl. (mmag) | Section | Figures |
|---|---|---|---|---|---|---|---|---|---|---|---|---|
| 02196 | 1973 | 02 21 20.25 | +57 06 27 | | | | | 4.507 | 52.16 | 1.4 | 8.1.2 | A.125-A.126 |
| 02231 | 2622 | 02 22 59.99 | +57 12 14 | | | | | 2.081 | 24.08 | 5.5 | 8.1.2 | A.127-A.128 |
| 02233 | 2141 | 02 21 49.63 | +57 01 36 | | | | | 0.209 | 2.42 | 54.6 | 8.1.2 | A.129-A.130 |
| 02234 | 2616 | 02 22 58.42 | +57 04 49 | | | | | 3.166 | 36.64 | 3.0 | 8.1.2 | A.131-A.132 |
| 02321 | 2342 | 02 22 13.13 | +57 01 22 | | | | | 2.678 | 30.99 | 4.5 | 8.1.2 | A.133-A.134 |
| 02351 | 2151 | 02 21 51.69 | +57 15 05 | | | | | 5.590 | 64.70 | 2.0 | 8.1.2 | A.135-A.136 |
| 02438 | 2563 | 02 22 46.99 | +56 58 06 | B2IIIe | | | | 3.664 | 42.41 | 5.0 | 8.1.2 | A.137-A.138 |
| 02465 | 2019 | 02 21 31.23 | +57 17 27 | | | | | 5.109 | 59.14 | 4.1 | 8.1.2 | A.139-A.140 |
| 02468 | 2611 | 02 22 57.42 | +57 00 32 | | | | | 4.703 | 54.43 | 3.1 | 8.1.2 | A.141-A.142 |
| 02475 | 2752 | 02 23 22.91 | +57 13 11 | | | | | 5.991 | 69.34 | 2.6 | 8.1.2 | A.143-A.144 |
| 02520 | 2725 | 02 23 18.49 | +57 15 33 | | | | | 1.811 | 20.96 | 4.6 | 8.1.2 | A.145-A.146 |
| 02542 | 2146 | 02 21 49.95 | +56 58 00 | | | | | 4.013 | 46.44 | 3.2 | 8.1.2 | A.147-A.148 |
| 02552 | 1898 | 02 21 10.93 | +56 59 32 | | | | | 3.648 | 42.22 | 3.1 | 8.1.2 | A.149-A.150 |
| 00003 | 2296 | 02 22 07.36 | +57 06 42 | B1III | 8.52 | 0.19 | 0.82 | | | | 8.1.3 | A.151 |
| 00007 | 2284 | 02 22 06.43 | +57 05 25 | B2III-IVe | 9.65 | 0.28 | 0.88 | | | | 8.1.3 | A.152 |
| 00009 | 2088 | 02 21 43.38 | +57 07 33 | B1IIIe | 9.44 | 0.20 | 0.94 | | | | 8.1.3 | A.153 |
| 00015 | 2165 | 02 21 52.92 | +57 09 59 | B2Ve | 10.04 | 0.20 | 0.83 | | | | 8.1.3 | A.154 |
| 00037 | 2085 | 02 21 42.92 | +57 05 30 | B3 | 11.22 | 0.25 | 0.74 | | | | 8.1.3 | A.155 |
| 00059 | 1977 | 02 21 24.91 | +57 11 53 | B2 | 12.24 | 0.29 | 0.56 | | | | 8.1.3 | A.156 |
| 02431 | 2402 | 02 22 22.76 | +57 17 04 | B1III | | | | | | | 8.1.3 | A.157 |
| 02434 | 1926 | 02 21 18.09 | +57 18 22 | B1IIIe | | | | | | | 8.1.3 | A.158 |
| 02447 | 2759 | 02 23 25.00 | +57 19 04 | B2 | | | | | | | 8.1.3 | A.159 |
| 00187 | 1970 | 02 21 23.66 | +57 06 30 | | 14.49 | 0.64 | -0.39 | 1.888 | 21.85 | 2.7 | 8.2.1 | A.160-A.161 |
| 00236 | 2500 | 02 22 37.92 | +57 03 11 | | 14.88 | 0.64 | -0.35 | 0.229 | 2.65 | 17.5 | 8.2.1 | A.162-A.163 |
| 00267 | 2544 | 02 22 43.97 | +57 03 54 | | 15.02 | 0.53 | -0.41 | 0.941 | 10.89 | 4.4 | 8.2.1 | A.164-A.165 |
| 00298 | 2430 | 02 22 27.71 | +57 13 32 | | 15.14 | 0.49 | -0.53 | 3.904 | 45.19 | 2.6 | 8.2.1 | A.166-A.167 |
| 00315 | 2306 | 02 22 09.40 | +57 04 21 | | 15.04 | 0.34 | -0.48 | 35.337 | 408.99 | 3.3 | 8.2.1 | A.168-A.169 |
| 00322 | 2243 | 02 23 03.02 | +57 11 19 | | 15.33 | 0.48 | -0.57 | 35.128 | 406.58 | 4.5 | 8.2.1 | A.170-A.171 |
| 00342 | 2416 | 02 22 24.16 | +57 05 48 | A3 | 15.27 | 0.36 | -0.49 | 15.298 | 177.06 | 1.6 | 8.2.1 | A.172-A.173 |
| 00344 | 2454 | 02 22 29.93 | +57 04 34 | | 15.47 | 0.51 | -0.67 | 0.840 | 9.73 | 7.6 | 8.2.1 | A.174-A.175 |
| 00364 | 2081 | 02 21 42.53 | +57 11 17 | | 15.37 | 0.43 | -0.56 | 35.754 | 413.82 | 2.4 | 8.2.1 | A.176-A.177 |
| 00370 | 2461 | 02 22 32.43 | +57 05 49 | | 15.50 | 0.45 | -0.52 | 32.249 | 373.25 | 1.6 | 8.2.1 | A.178-A.179 |
| 00371 | 4990 | 02 21 57.60 | +57 06 16 | | 15.64 | 0.58 | -0.41 | 1.083 | 12.53 | 13.6 | 8.2.1 | A.180-A.181 |
| 00381 | 2476 | 02 22 34.55 | +57 09 47 | | 15.47 | 0.42 | -0.55 | 36.614 | 423.78 | 1.8 | 8.2.1 | A.182-A.183 |
| 00384 | 2220/3379 | 02 21 59.96 | +57 05 10 | | 15.49 | 0.44 | -0.57 | 35.236 | 407.82 | 2.8 | 8.2.1 | A.184-A.185 |
| 00388 | 5155 | 02 22 06.79 | +57 04 29 | | 15.56 | 0.46 | -0.47 | 15.584 | 180.37 | 3.6 | 8.2.1 | A.186-A.187 |
| 00419 | 5168 | 02 21 50.35 | +57 04 36 | | 15.68 | 0.50 | -0.59 | 23.346 | 270.21 | 5.5 | 8.2.1 | A.188-A.189 |
| 00471 | 5151 | 02 22 16.04 | +57 04 51 | | 15.82 | 0.47 | -0.48 | 23.650 | 273.73 | 4.7 | 8.2.1 | A.190-A.191 |
| 00540 | 3276 | 02 22 08.99 | +57 09 55 | | 16.07 | 0.54 | -0.29 | 22.962 | 265.76 | 9.6 | 8.2.1 | A.192-A.193 |
| 00586 | 4830 | 02 22 38.31 | +57 09 03 | | 16.61 | 0.70 | -0.35 | 0.728 | 8.43 | 8.6 | 8.2.1 | A.194-A.195 |
| 02249 | 7057 | 02 22 58.89 | +57 03 45 | | | | | 8.411 | 97.35 | 3.7 | 8.2.1 | A.196-A.197 |
| 00332 | 2445 | 02 22 29.55 | +57 09 37 | | 15.18 | 0.35 | -0.57 | 42.204 | 488.47 | 1.3 | 8.2.2 | A.198-A.199 |
| 00376 | 5172 | 02 21 52.42 | +57 02 15 | | | | | 10.585 | 122.51 | 9.1 | 8.2.2 | A.200-A.201 |
| 00389 | 2062 | 02 21 38.00 | +57 02 17 | | | | | 15.710 | 181.83 | 2.9 | 8.2.2 | A.202-A.203 |
| 00412 | 3344 | 02 21 50.11 | +57 06 49 | | 15.72 | 0.53 | -0.53 | 23.169 | 268.16 | 2.3 | 8.2.2 | A.204-A.205 |
| 00417 | 2206 | 02 21 58.56 | +57 07 11 | A7 | 15.58 | 0.43 | -0.49 | 23.440 | 271.30 | 4.5 | 8.2.2 | A.206-A.207 |
| 00421 | 2030 | 02 21 32.95 | +57 04 35 | | 15.69 | 0.48 | -0.53 | 18.238 | 211.09 | 4.7 | 8.2.2 | A.208-A.209 |
| 00428 | 5127 | 02 22 34.24 | +57 03 11 | | 15.70 | 0.45 | -0.52 | 20.088 | 232.50 | 2.0 | 8.2.2 | A.210-A.211 |
| 00438 | | 02 22 34.75 | +57 12 40 | | 15.75 | 0.45 | -0.55 | 22.420 | 259.50 | 3.5 | 8.2.2 | A.212-A.213 |
| 00443 | 4832 | 02 22 30.65 | +57 09 36 | | 15.77 | 0.46 | -0.45 | 49.457 | 572.42 | 1.9 | 8.2.2 | A.214-A.215 |
| 00450 | 2339 | 02 22 13.12 | +57 07 54 | | 15.76 | 0.44 | -0.61 | 48.455 | 560.82 | 2.2 | 8.2.2 | A.216-A.217 |
| 00462 | 4713 | 02 21 36.69 | +57 12 59 | | 15.90 | 0.50 | -0.53 | 20.528 | 237.59 | 3.7 | 8.2.2 | A.218-A.219 |
| 00497 | 4829 | 02 22 37.34 | +57 08 19 | | 15.92 | 0.46 | -0.48 | 41.166 | 476.46 | 2.1 | 8.2.2 | A.220-A.221 |
| 00503 | 3363 | 02 21 46.70 | +57 06 32 | | 15.96 | 0.52 | -0.51 | 22.753 | 263.35 | 2.6 | 8.2.2 | A.222-A.223 |
| 00509 | 3212 | 02 22 11.13 | +57 05 46 | | 15.98 | 0.49 | -0.48 | 22.379 | 259.01 | 2.5 | 8.2.2 | A.224-A.225 |
| 00527 | 4819 | 02 22 42.59 | +57 09 54 | | 15.93 | 0.37 | -0.58 | 20.696 | 239.54 | 2.1 | 8.2.2 | A.226-A.227 |
| 00570 | 5160 | 02 22 02.62 | +57 03 29 | | 16.31 | 0.55 | -0.41 | 1.577 | 18.26 | 22.7 | 8.2.2 | A.228-A.229 |
| 00577 | 5183 | 02 21 36.26 | +57 04 27 | | 16.50 | 0.64 | -0.45 | 8.258 | 95.57 | 4.9 | 8.2.2 | A.230-A.231 |
| 00744 | 4978 | 02 22 22.90 | +57 05 04 | | 16.57 | 0.39 | -0.56 | 15.022 | 173.87 | 3.7 | 8.2.2 | A.232-A.233 |
| 02205 | 1929 | 02 21 18.49 | +57 09 26 | | | | | 35.590 | 411.92 | 4.4 | 8.2.2 | A.234-A.235 |
| 02376 | 2210 | 02 21 59.76 | +57 15 25 | | | | | 9.172 | 106.16 | 5.2 | 8.2.2 | A.236-A.237 |
| 02414 | | 02 21 37.41 | +57 14 12 | | 17.16 | 0.59 | -0.45 | 3.914 | 45.26 | 16.3 | 8.2.2 | A.238-A.239 |
| 02727 | 2671 | 02 23 08.19 | +57 01 02 | | | | | 35.767 | 413.97 | 3.1 | 8.2.2 | A.240-A.241 |
| 02740 | 2703 | 02 23 14.41 | +57 11 40 | | | | | 1.569 | 18.16 | 17.8 | 8.2.2 | A.242-A.243 |
| 02833 | 4398 | 02 21 55.01 | +57 18 42 | | | | | 46.878 | 542.57 | 4.4 | 8.2.2 | A.244-A.245 |
| 00521 | 3267 | 02 22 15.62 | +57 09 45 | | 16.02 | 0.55 | -0.05 | 1.980 | 22.91 | 19.0 | 8.3 | A.246-A.247 |

**Table A.1.** Continued.

| Star ID | WEBDA | $\alpha$ (h m s) | $\delta$ (° ′ ″) | SpT | $V$ | $B_2 - V_1$ | $B_2 - U$ | Frequency ($d^{-1}$) | ($\mu$Hz) | Ampl. (mmag) | Section | Figures |
|---|---|---|---|---|---|---|---|---|---|---|---|---|
| 00565 | 5145 | 02 22 21.64 | +57 04 00 | | 16.71 | 0.91 | -0.63 | 2.291 | 26.52 | 13.0 | 8.3 | A.248-A.249 |
| 00631 | 6911 | 02 22 26.75 | +57 11 29 | | 16.97 | 0.83 | -0.72 | 2.367 | 27.40 | 15.1 | 8.3 | A.250-A.251 |
| 00681 | 4839 | 02 22 15.05 | +57 10 58 | | 17.18 | 0.94 | -0.68 | 1.818 | 21.04 | 28.6 | 8.3 | A.252-A.253 |
| 00719 | | 02 22 12.60 | +57 06 03 | | 17.10 | 0.88 | -0.42 | 0.862 | 9.98 | 30.1 | 8.3 | A.254-A.255 |
| 00726 | | 02 22 31.28 | +57 04 54 | | 17.26 | 0.89 | -0.53 | 2.238 | 25.90 | 17.3 | 8.3 | A.256-A.257 |
| 00765 | 2359 | 02 22 15.37 | +57 08 26 | B9 | 17.35 | 0.98 | -0.60 | 1.872 | 21.67 | 19.2 | 8.3 | A.258-A.259 |
| 00805 | | 02 22 10.49 | +57 03 55 | | 17.41 | 0.50 | -0.18 | 2.420 | 28.01 | 18.8 | 8.3 | A.260-A.261 |
| 00816 | 6870 | 02 22 19.69 | +57 09 53 | | 17.49 | 0.89 | -0.60 | 0.779 | 9.02 | 42.3 | 8.3 | A.262-A.263 |
| 00827 | | 02 21 39.01 | +57 11 08 | | 17.48 | 0.88 | -0.60 | 0.239 | 2.77 | 30.6 | 8.3 | A.264-A.265 |
| 01098 | | 02 21 54.01 | +57 06 31 | | 18.40 | 1.01 | -0.76 | 6.682 | 77.34 | 47.2 | 8.3 | A.266-A.267 |
| 02334 | 5329 | 02 22 21.68 | +57 01 31 | | | | | 9.524 | 110.23 | 10.2 | 8.3 | A.268-A.269 |
| 02478 | 2681/5292 | 02 23 09.25 | +56 59 48 | | | | | 0.058 | 0.67 | 12.9 | 8.3 | A.270-A.271 |
| 02507 | 1964 | 02 21 22.73 | +57 01 24 | | | | | 6.953 | 80.48 | 8.2 | 8.3 | A.272-A.273 |
| 02513 | 2741 | 02 23 19.18 | +56 56 10 | | | | | 1.181 | 13.67 | 14.6 | 8.3 | A.274-A.275 |
| 02534 | 2670/4943 | 02 23 08.62 | +57 07 07 | | | | | 0.032 | 0.37 | 30.9 | 8.3 | A.276-A.277 |
| 02543 | 2779 | 02 23 29.14 | +57 14 11 | | | | | 29.031 | 336.01 | 2.7 | 8.3 | A.278-A.279 |
| 02909 | 1866 | 02 21 06.16 | +57 11 37 | | | | | 1.591 | 18.41 | 19.2 | 8.3 | A.280-A.281 |
| 02953 | 5101 | 02 23 18.10 | +57 02 30 | | | | | 0.103 | 1.19 | 54.2 | 8.3 | A.282-A.283 |
| 03114 | 5205 | 02 21 11.64 | +57 03 59 | | | | | 2.189 | 25.33 | 27.6 | 8.3 | A.284-A.285 |
| 00091 | 2107 | 02 21 45.54 | +57 05 30 | | 13.88 | 1.16 | -1.29 | 0.824 | 9.54 | 2.1 | 8.3 | A.286-A.287 |
| 00507 | 4947 | 02 22 54.02 | +57 07 34 | | 16.57 | 0.93 | -0.71 | 0.331 | 3.83 | 35.6 | 8.3 | A.288-A.289 |
| 00549 | 3242 | 02 22 17.91 | +57 06 48 | | 16.70 | 0.96 | -0.54 | 2.707 | 31.34 | 11.1 | 8.3 | A.290-A.291 |
| 00660 | 3209 | 02 22 16.19 | +57 05 39 | | 17.03 | 0.87 | -0.61 | 0.570 | 6.60 | 20.8 | 8.3 | A.292-A.293 |
| 00667 | 5176 | 02 21 45.55 | +57 02 25 | | 16.76 | -0.51 | 0.01 | 2.113 | 24.46 | 32.0 | 8.3 | A.294-A.295 |
| 00706 | | 02 21 43.93 | +57 07 52 | | 17.06 | 0.73 | -0.11 | 5.084 | 58.84 | 126.7 | 8.3 | A.296-A.297 |
| 00710 | 5175 | 02 21 46.45 | +57 02 44 | | 17.36 | 1.08 | -0.73 | 2.594 | 30.02 | 15.5 | 8.3 | A.298-A.299 |
| 00738 | | 02 22 17.23 | +57 07 32 | | 16.50 | -0.07 | 3.19 | 3.013 | 34.87 | 58.5 | 8.3 | A.300-A.301 |
| 00834 | | 02 21 39.77 | +57 11 19 | | 17.80 | 1.02 | -0.67 | 8.081 | 93.53 | 12.1 | 8.3 | A.302-A.303 |
| 00907 | | 02 22 07.42 | +57 13 42 | | 17.70 | 0.69 | -0.67 | 5.478 | 63.41 | 50.3 | 8.3 | A.304-A.305 |
| 00940 | 4716 | 02 21 25.34 | +57 11 21 | | 18.05 | 1.06 | -0.90 | 3.711 | 42.95 | 53.4 | 8.3 | A.306-A.307 |
| 02476 | 2791 | 02 23 30.41 | +57 02 42 | | | | | 5.645 | 65.33 | 1.9 | 8.3 | A.308-A.309 |
| 02480 | 1900 | 02 21 12.21 | +57 12 47 | | | | | 0.329 | 3.80 | 10.4 | 8.3 | A.310-A.311 |
| 02684 | 1915 | 02 21 16.13 | +57 17 28 | | | | | 24.667 | 285.50 | 6.0 | 8.3 | A.312-A.313 |
| 02826 | 4728 | 02 21 00.65 | +57 13 33 | | | | | 0.409 | 4.74 | 38.9 | 8.3 | A.314-A.315 |
| 02860 | 4471 | 02 23 23.05 | +57 14 42 | | | | | 13.464 | 155.83 | 4.6 | 8.3 | A.316-A.317 |
| 02865 | 5226 | 02 21 03.34 | +57 02 54 | | | | | 35.193 | 407.33 | 4.4 | 8.3 | A.318-A.319 |
| 02900 | 4489 | 02 22 53.26 | +57 16 55 | | | | | 11.339 | 131.24 | 5.8 | 8.3 | A.320-A.321 |
| 03122 | 5466 | 02 22 32.16 | +56 58 13 | | | | | 12.034 | 139.29 | 51.6 | 8.3 | A.322-A.323 |
| 03146 | | 02 21 32.62 | +57 00 32 | | | | | 2.698 | 31.23 | 25.8 | 8.3 | A.324-A.325 |
| 00001 | 2417 | 02 22 24.28 | +57 06 34 | M4Iab | | | | | | | 8.3 | A.326 |
| 00002 | 2227 | 02 22 00.56 | +57 08 41 | B2Ibp | | | | | | | 8.3 | A.327 |
| 02427 | 2691/4637 | 02 23 11.07 | +57 11 58 | M1Iab | | | | | | | 8.3 | A.328 |
| 02428 | 2758/4624 | 02 23 24.13 | +57 12 44 | M1Iab | | | | | | | 8.3 | A.329 |
| 03164 | 1818 | 02 20 59.63 | +57 09 31 | sdM1 | | | | | | | 8.3 | A.330 |
| 03165 | 2589 | 02 22 53.51 | +57 14 42 | A2Ia | | | | | | | 8.3 | A.331 |

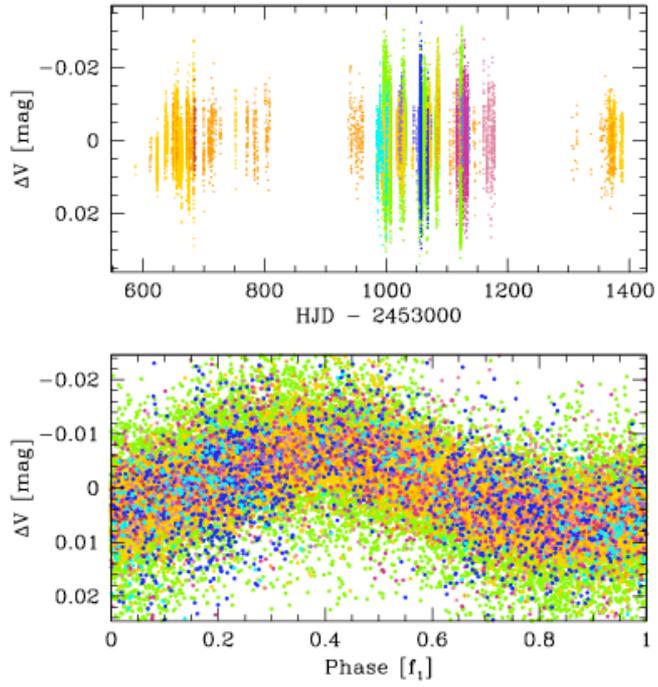

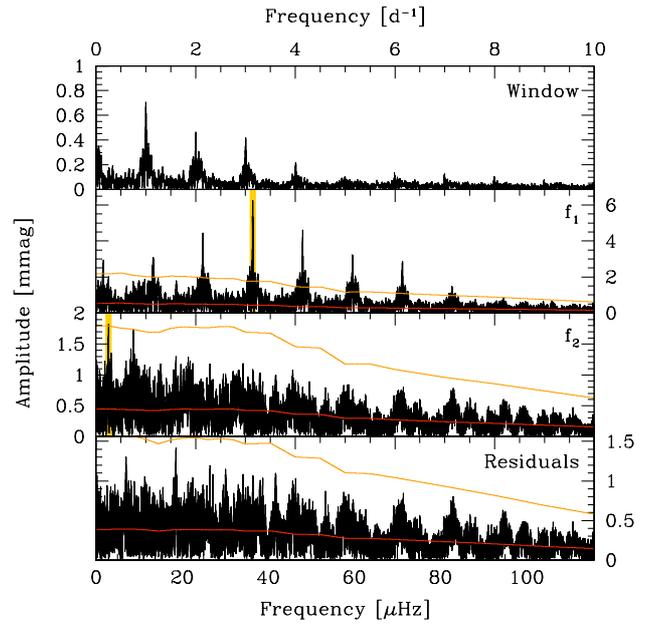

**Fig. A.2.** Same as Fig. 11, but for star 00004.

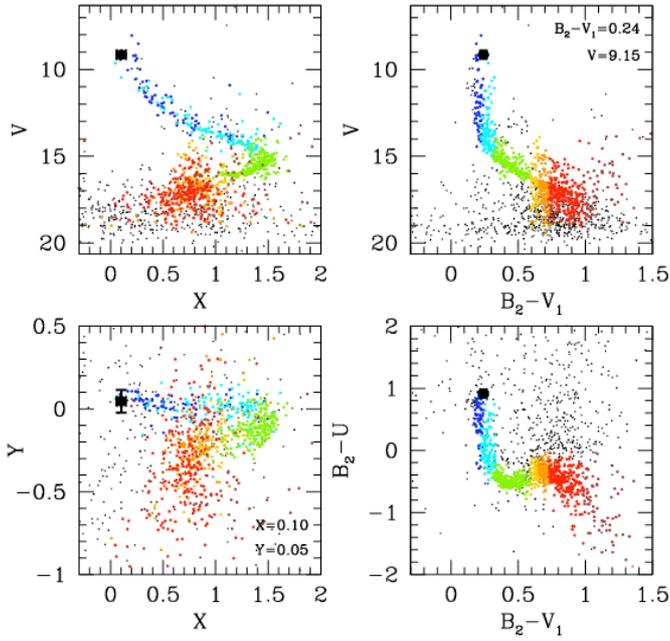

**Fig. A.1.** Same as Fig. 10, but for star 00004.